\documentclass[aps,prb,reprint,superscriptaddress,amsmath,amssymb]{revtex4-2}
\usepackage{graphicx}
\usepackage{bm}
\usepackage{amsmath,amssymb}
\usepackage{booktabs,array,longtable}
\usepackage{xcolor}
\usepackage[margin=0.72in]{geometry}
\usepackage[colorlinks=true,linkcolor=blue,filecolor=magenta,
urlcolor=blue]{hyperref}
\usepackage{multirow}

\newcommand{\SPC}{\mathrm{SPC}}
\newcommand{\KM}{\mathrm{KM}}
\newcommand{\se}{s_{\varepsilon}}
\newcommand{\sbond}{s_b}

\begin{document}

\title{Where electrons conduct and where to change it: gauge-invariant sensitivities and the networks of sensitive bonds}

\author{D. A. Drabold}
\email{drabold@ohio.edu}
\affiliation{Department of Physics and Astronomy, Ohio University, Athens, Ohio 45701, USA}
\author{C. Ugwumadu}
\affiliation{Theoretical Division (T-4), Los Alamos National Laboratory, Los Alamos, New Mexico 87545, USA}

\date{\today}

\begin{abstract}
Where is the electrical current carried in a disordered solid, and where would a local change alter it? The two questions differ. We answer the second exactly: the derivative of the Kubo--Greenwood conductivity with respect to the Hamiltonian is gauge invariant and gives the response to any local change. The construction applies to any Hamiltonian. We try it on amorphous silicon, with tight-binding cells of up to 10\,000 atoms and density-functional cells of up to 512 atoms, and compare it with three ways of allocating the conductivity: the space-projected conductivity (SPC), KuboMap, and the local density of states (LDOS). KuboMap follows the LDOS closely (median Spearman correlation 0.97). The sensitivity does not. It ranks the response to an on-site shift accurately and the response to removing an atom poorly, and it transfers between Hamiltonians less well than the other maps. The most sensitive bonds form connected clusters. In three large networks they are tree-like filaments of 24 to 38 atoms in the valence tail and looped networks deeper in the gap; the excess over an LDOS-weighted random selection reaches 48 standard deviations and appears only when the broadening spans about five level spacings. The filament overlaps the tail eigenstate with the most weight on it, and one of the leading conduction channels, far more than a random cluster does, but no single state or channel explains it. The networks are largely lost in thermal snapshots at 300 K. Where the conductivity is carried and where it responds are different problems: local transport maps are useful for the first and limited for the second.
\end{abstract}

\maketitle

\section{Introduction}
Where is the electrical current carried in a disordered solid, and where would a local change alter it? The two questions differ. The Kubo--Greenwood formula gives one number for the conductivity $\sigma$~\cite{Kubo1957,Greenwood1958,Allen2006transport} and is widely used for amorphous silicon and other disordered systems~\cite{Allen1987,Abtew2007sigma,Zhang2010PRL,Calderin2017}. Several constructions indicate where in real space that number comes from: the space-projected conductivity (SPC)~\cite{Kiran2018SPC_1,Subedi2020SPC_2}, pair-overlap schemes in the spirit of Hindley and Mott~\cite{Hindley1970,Mott1969,Nepal2025N2,Hussein2026TAHM}, and the recent KuboMap~\cite{Ugwumadu2026KuboMap}, a non-negative density built from products of the probability densities of pairs of eigenstates. All of these are projections, and a projection is not unique. The total is fixed; its division among atoms is not. Any construction that uses individual eigenvectors has a further ambiguity in degenerate subspaces, the ``gauge''~\cite{Zhang2010degenerate}.

The second question has an exact, gauge-free answer: the derivative of $\sigma$ with respect to the Hamiltonian. In a companion paper~\cite{Drabold2026SPTC} we found the same pattern for heat. The site-projected thermal conductivity (SPTC)~\cite{Gautam2025SPTC_1,Ugwumadu2026SPTC_2,Ugwumadu2025galactic,Ugwumadu2026SCACS} depends moderately on the arbitrary division of the potential energy, while the response of the thermal conductivity $\kappa$ to a change of mass or force constant is gauge invariant, because $\kappa$ is. This paper is the electronic analogue. The ideas apply to any Hamiltonian; we try them on amorphous silicon. We ask how the two answers are related and how well each projection recovers the atoms to which $\sigma$ responds.

This paper makes two contributions. The first is a method: the derivative $G=\partial\sigma/\partial H$ is gauge invariant, obeys two exact sum rules, contains the space-projected conductivity as one of the three terms of its bond part, and ranks the exact response to a finite on-site shift at Spearman 0.97--1.00. The second is a finding: the most sensitive bonds are not scattered. In three large networks they form tree-like filaments of 24 to 38 atoms in the valence tail and looped networks deeper in the gap, tied to tail eigenstates and to the leading conduction channels, once the broadening spans about five level spacings. The allocation maps, SPC, KuboMap, and the LDOS, locate this response only in part: KuboMap follows the LDOS closely, and no map predicts the response to removing an atom.

The question matters most in disordered solids, where $\sigma$ is controlled by features beyond the content of the density of states: tail states, coordination defects, and networks of strained bonds~\cite{Anderson1958,Mott1969,MottDavis1979,Thouless1974,Edwards1972localization,Abrahams1979scaling,Dong1998BandTail,Fedders1998,Pan2008,Drabold2011Urbach}. Where carriers are localized, conduction proceeds by phonon-assisted hopping~\cite{Miller1960hopping,Ambegaokar1971hopping,Austin1969polarons,MottDavis1979,Li2003hopping}, and solving the time-dependent Schr\"odinger equation is an alternative route~\cite{DraboldLi2002TDSE,Li2003hopping}. The band-conduction formula used here does not describe that regime. Related atom-resolved current and pathway analyses exist for tight-binding and molecular-junction transport~\cite{Todorov2002,Solomon2010}.

We compute $\sigma$ for localized-orbital Hamiltonians and the atomic positions, compare the three atom-resolved partitions, and differentiate $\sigma$ with respect to local perturbations of $H$. The derivatives define gauge-invariant on-site, bond-strength, and position sensitivities.

\section{Theory}
\subsection{Conductivity and partitions}
We work in a real, orthogonal localized basis. The one-electron Hamiltonian is a real symmetric matrix, \(H=H^{\mathsf T}\), with normalized eigenvectors collected in \(E\):
\begin{equation}
 HE=E\varepsilon,\quad E^{\mathsf T}E=I,\quad
 \varepsilon=\operatorname{diag}(\epsilon_1,\ldots,\epsilon_M).
\label{eq:app_eigen}
\end{equation}
An orbital index \(a\) belongs to atom \(x(a)\), and \(\Pi_x\) denotes the diagonal projector onto the orbitals centered on atom \(x\). The atomic projection of eigenstate \(i\) is
\begin{equation}
 n_i(x)=E_i^{\mathsf T}\Pi_xE_i
       =\sum_{a\in x}E_{ai}^2,
 \qquad \sum_x n_i(x)=1.
 \label{eq:n_i}
\end{equation}

For a periodic structure, the oriented minimum-image displacement is
\begin{equation}
 R^\alpha_{ab}
 =\bigl[r_{x(b)}-r_{x(a)}\bigr]^\alpha_{\mathrm{MI}},
 \qquad R^\alpha_{ba}=-R^\alpha_{ab}.
\label{eq:app_R}
\end{equation}
We define the velocity-related matrix
\begin{equation}
 A^\alpha=H\circ R^\alpha,
\end{equation}
where \(\circ\) denotes the elementwise (Hadamard) product, \((X\circ Y)_{ab}=X_{ab}Y_{ab}\), and \(R^\alpha\) is the displacement matrix of Eq.~\eqref{eq:app_R}. Since \(H\) is symmetric and \(R^\alpha\) is antisymmetric, \(A^\alpha\) is antisymmetric. Its representation in the eigenstate basis,
\begin{equation}
 P^\alpha=E^{\mathsf T}A^\alpha E,
\end{equation}
is therefore also antisymmetric, so that \(P^\alpha_{ji}=-P^\alpha_{ij}\) and \((P^\alpha_{ij})^2\geq0\). In this representation, the velocity matrix elements are proportional to \(A^\alpha\), with \(v^\alpha=iA^\alpha/\hbar\).

The Fermi--Dirac occupation and Gaussian energy-broadening kernel are
\begin{equation}\label{eq:app_fermi_gaussian}
\begin{aligned}
f(\epsilon;\mu, T)
  &= \frac{1}{\exp\!\left(\frac{\epsilon-\mu}{k_B T}\right)+1},\\
\delta_\eta(u)
  &= \frac{1}{\sqrt{2\pi}\,\eta}\exp\!\left(-\frac{u^2}{2\eta^2}\right).
\end{aligned}
\end{equation}

The Kubo--Greenwood weight is defined as
\begin{equation}
\begin{aligned}
 W_{ij} &=~ w(\epsilon_i,\epsilon_j)\\
 &=\frac{f(\epsilon_i;\mu,T)-f(\epsilon_j;\mu,T)}
 {\epsilon_j-\epsilon_i}\,
 \delta_\eta(\epsilon_i-\epsilon_j).
\label{eq:weight}
\end{aligned}
\end{equation}
For \(i=j\), the occupation ratio is evaluated by continuity:
\begin{equation}
 W_{ii}=-f'(\epsilon_i;\mu,T)\delta_\eta(0),
 \qquad
 \delta_\eta(0)=\frac{1}{\sqrt{2\pi}\eta}.
\label{eq:weightdiag}
\end{equation}
For the Fermi--Dirac distribution, \(W_{ij}=W_{ji}\geq0\).

The isotropic scalar conductivity is then
\begin{equation}
 \sigma=\frac{C}{3}\sum_{\alpha}\sum_{ij}
 W_{ij}(P^\alpha_{ij})^2,
 \qquad
 C=\frac{\pi e^2s}{\hbar\Omega},
\label{eq:sigma}
\end{equation}
where \(e\) is the elementary charge, \(s\) is the spin multiplicity, \(\hbar\) is the reduced Planck constant, and \(\Omega\) is the cell volume. The index \(\alpha\in\{x,y,z\}\) denotes the Cartesian direction, while \(i\) and \(j\) label the electronic eigenstates. The tensor conductivity and the complex-Hermitian case follow by the obvious replacements; we use the isotropic scalar form throughout.

To examine the spatial distribution of the electronic conductivity, we consider two atom-resolved partitions of the total Kubo--Greenwood conductivity. The first is the space-projected conductivity (SPC) \cite{Kiran2018SPC_1,Subedi2020SPC_2}, built from a nonlocal conduction matrix \cite{Baranger1989},
\begin{align}
 \sigma_{\mathrm{spc}}(x)
 &=\frac{C}{3}\sum_\alpha\sum_{a\in x,b}
 A^\alpha_{ab}B^\alpha_{ab},\\
 B^\alpha&=E(W\circ P^\alpha)E^{\mathsf T},
\label{eq:partitions_spc}
\end{align}
where \(B^\alpha\) is constructed by transforming the elementwise-weighted current-related matrix \(W\circ P^\alpha\) back into the localized orbital basis. Because the weights \(W_{ij}\) depend only on the eigenvalues [Eq. \eqref{eq:weight}], \(B^\alpha\) is invariant under orthogonal rotations within exactly degenerate eigenspaces. The SPC partition is therefore gauge invariant. It is not positive: individual atomic contributions can be negative, although their sum is the non-negative total conductivity.

As a second partition, we consider the atom-resolved KuboMap,
\begin{equation}
\begin{aligned}
 \rho_{\mathrm{km}}(x)
 &=\frac{C}{3}\sum_{\alpha}\sum_{ij}
 W_{ij}\,(P^\alpha_{ij})^2\,\pi_{ij}(x),\\
 \pi_{ij}(x)&=\frac{n_i(x)\,n_j(x)}{\sum_y n_i(y)\,n_j(y)}.
\end{aligned}
\label{eq:partitions_km}
\end{equation}
Here, the weight \(\pi_{ij}(x)\), built from the product \(n_i(x)n_j(x)\), distributes the conductivity contribution associated with each eigenstate pair \((i,j)\) among the atoms according to their spatial populations. Unlike SPC, this partition is generally gauge dependent, since the individual atomic populations \(n_i(x)\) can change under rotations within degenerate eigenspaces. We address this limitation through a spectral-subspace formulation of KuboMap, introduced in Appendix~\ref{app:binned}, which restores invariance under such rotations. The atom-resolved formulation of KuboMap used here also provides a discrete analogue of the grid-based KuboMap of Ref.~\cite{Ugwumadu2026KuboMap}. We discuss the similarities and differences between the grid- and atom-resolved variants of KuboMap in Sec.~S6 of the Supplemental Material.

Provided the denominator is nonzero for every contributing state pair, both partitions recover the total conductivity:
\begin{equation}
 \sum_x\sigma_{\mathrm{spc}}(x)
 =\sum_x\rho_{\mathrm{km}}(x)
 =\sigma.
\end{equation}

As an independent reference for the spatial distribution of electronic states, we also consider the local density of states (LDOS), defined here using Gaussian energy broadening:
\begin{equation}
 \rho_\mathrm{ldos}(x;\mu,\eta)
 =\sum_i n_i(x)\,
 \delta_\eta(\epsilon_i-\mu),
\label{eq:partitions_ldos}
\end{equation}
where \(\mu\) is the chemical potential and \(\delta_\eta\) is the normalized Gaussian kernel defined in Eq.~\eqref{eq:app_fermi_gaussian}. Unlike the conductivity partitions, the LDOS is not a decomposition of the conductivity; rather, it provides a gauge-invariant measure of the local electronic spectral weight near the chemical potential. Its gauge invariance follows because the spectral weight is unchanged by rotations within degenerate eigenspaces.

\subsection{Gauge-free sensitivities}

Because the total conductivity \(\sigma\) is a basis-independent scalar, its derivative with respect to a physical parameter of the Hamiltonian $H$ is also basis independent, provided the basis transformation is independent of that parameter (or its parameter dependence is treated consistently). In what follows, we define $G=\partial\sigma/\partial H$, and differentiate with respect to different components of $H$.

\subsubsection{local energy and bond derivatives}
First we define a shift $\varepsilon_x$ of the on-site energies of atom $x$ as
\begin{equation}
 \se(x)=\frac{\partial\sigma}{\partial\varepsilon_x}=\sum_{a\in x}G_{aa},
\label{eq:sens-energy}
\end{equation}
a scaling $\lambda_{xy}$ of the blocks $H_{xy},H_{yx}$ as
\begin{equation}
 \sbond(x,y)=\frac{\partial\sigma}{\partial\ln\lambda_{xy}}=2\sum_{a\in x,b\in y}G_{ab}H_{ab}
\label{eq:sens-bond}
\end{equation}
Differentiating Eq.~\eqref{eq:sigma} with divided differences (the Daleckii--Krein formula) gives \cite{Daleckii1965integration,Bhatia1997matrix}
\begin{equation}
 G=\operatorname{sym}\!\left(E\widetilde{G}E^{\mathsf T}+G^{(A)}\right),
\label{eq:gradient}
\end{equation}
where $\operatorname{sym}(X)=\tfrac{1}{2}(X+X^{\mathsf T})$. The matrix $\widetilde G$ in the eigenbasis and the explicit term $G^{(A)}$ are given in Eqs.~\eqref{eq:app_GA}--\eqref{eq:app_Goffdiag}. $\widetilde G_{ii}$ is the eigenvalue response (via the Hellmann--Feynman theorem~\cite{Feynman1939forces} $\partial\epsilon_i/\partial\ln\lambda_{xy}=2\sum H_{ab}E_{ai}E_{bi}$), the off-diagonal part is the eigenvector response, and $G^{(A)}$ is the explicit dependence through the velocity operator. The first-variation derivation used here is given in Appendix~\ref{app:gradient}, its numerical treatment and exactly degenerate case are described in Appendix~\ref{app:degeneracy}.

Derived from continuous symmetries of the Kubo--Greenwood functional [see Appendix~\ref{app:sumrules}], two exact sum rules hold:
\begin{equation}
\begin{aligned}
    &\sum_x\se(x)=-\partial_\mu\sigma,\\
    &\sum_a\epsilon_aG_{aa}+\sum_{a\ne b}H_{ab}G_{ab}+\mu\partial_\mu\sigma+\eta\partial_\eta\sigma+T\partial_T\sigma=0.
    \label{eq:sumrules}
\end{aligned}  
\end{equation}
The second follows because $\sigma$ is invariant under $H,\mu,\eta,k_BT\mapsto\lambda(\cdot)$: $W\mapsto W/\lambda^2$ and $P^2\mapsto\lambda^2P^2$. 

SPC turns out to be one term of the gradient. Contracting $G^{(A)}$ with $H$ gives the explicit term of the bond sensitivity, twice the bond-resolved SPC,
\begin{equation}
 \sbond(x,y)=2\sigma_{\SPC}(x,y)+s_b^{\rm eig}(x,y)+s_b^{\rm vec}(x,y).
\label{eq:bonddecomp}
\end{equation}
Equation~\eqref{eq:bonddecomp} was verified numerically to $2\times10^{-16}$. SPC records how $\sigma$ changes because the velocity operator changes; the response of the levels and states is not in it. The bond-scaling perturbation and the decomposition into SPC, eigenvalue, and eigenvector contributions are derived in Appendix~\ref{app:localderivatives}.

\subsubsection{Position derivative}
Consider two atoms $I$ and $J$ with $d=X_J-X_I$ and an orbital block $T_{oo'}(d)=H_{Io,Jo'}$.  Translational invariance gives $\partial\sigma/\partial X_J=-\partial\sigma/\partial X_I=\partial\sigma/\partial d$.  Two contributions arise: the change of the hopping block, which enters through the full gradient $G$, and the change of the displacement in the velocity operator $A^\beta=H\circ R^\beta$.  Differentiating both gives
\begin{equation}
 \begin{aligned}
 \frac{\partial\sigma}{\partial d^\alpha}={}&
 \sum_{oo'}(G_{Io,Jo'}+G_{Jo',Io})\frac{\partial T_{oo'}}{\partial d^\alpha}\\
 &+\frac{4C}{3}\sum_{oo'}B^\alpha_{Io,Jo'}H_{Io,Jo'}.
 \end{aligned}
\label{eq:app_positionfull}
\end{equation}

With $\widehat d=d/|d|$, the radial and transverse responses are
\begin{equation}
 s_{\parallel}=\widehat d\cdot\frac{\partial\sigma}{\partial d},\quad
 \bm s_{\perp}=\left(I-\widehat d\widehat d^{\mathsf T}\right)\frac{\partial\sigma}{\partial d},
 \quad s_{\perp}=|\bm s_{\perp}|.
\label{eq:app_radialtransverse}
\end{equation}
Thus $s_{\parallel}$ is the first-order response to bond stretching, whereas $\bm s_\perp$ is the response to rotating the bond at fixed length.  In a two-centre model the first term is radial, so a transverse response remains only through the direction-dependent current operator.

\subsubsection{Other local derivatives and remarks}

SPC and KuboMap are spatial partitions of the already computed conductivity: both sum to $\sigma$, but neither is a derivative of $\sigma$ with respect to a physical perturbation.  In contrast, $\se(x)$, $\sbond(x,y)$, and the position derivative are response functions.  They quantify how $\sigma$ changes when a local on-site energy, hopping strength, or atomic position is infinitesimally modified.

More generally, any derivative of $\sigma$ with respect to a local parameter of $H$ is gauge invariant: the on-site energy of one orbital; the strength of one hopping element; an atomic or pair position; local volume or strain; and, at each $\mu$, the derivative of the Mott thermopower $\propto\partial\ln\sigma/\partial\mu$. Derivatives with respect to $\mu$, $\eta$, and $T$ are global and enter through the sum rules [Eq. \eqref{eq:sumrules}]. The maps answer different questions: $\se$ identifies where to place disorder, a dopant, or a different species; $\sbond$ identifies responsive bonds; and position derivatives identify how to move atoms.

KuboMap built from individual eigenstates depends on the choice of eigenvectors in degenerate subspaces. Binning the spectrum into energy windows and using subspace invariants removes this dependence within each bin, at the price of approximate closure and a dependence on the bin edges. We give the construction and its proof of invariance in Appendix~\ref{app:binned}, and the numerical tests in Sec.~IV~E and Sec.~S8.

\begin{table}[t]
\centering
\footnotesize
\setlength{\tabcolsep}{4pt}
\renewcommand{\arraystretch}{1.15}
\caption{Coordination-defect statistics using a
2.8~\AA\ neighbour cutoff. Defective sites are
non-fourfold atoms. Ranges indicate variation
within a model family. R10 and R2--R4 are the original structures; the ACE-relaxed R2--R4 of the thermal study have the same counts.}
\label{tab:si_models}
\begin{ruledtabular}
\begin{tabular}{lccc}
Model & 3-/5-/6-fold & Defective sites & Defects (\%) \\
\hline
R10            & 0 / 0 / 0   & 0     & 0 \\
R2             & 10 / 0 / 0  & 10    & 1.0 \\
R3             & 5 / 3 / 0   & 8     & 0.8 \\
R4             & 3 / 3 / 0   & 6     & 0.6 \\
aSi-216-A--D   & varies     & 2---14 & 0.93---6.48 \\
aSi-512-A--C   & varies     & 6---24 & 1.17---4.69 \\
DTW-512        & 0 / 0 / 0   & 0     & 0 \\
DTW-4096-U       & 0 / 10 / 0  & 10    & 0.24 \\
DTW-4096 & 5 / 23 / 0  & 28    & 0.68 \\
BM-10K         & 16 / 36 / 4 & 56    & 0.56 \\
\end{tabular}
\end{ruledtabular}
\end{table}

\section{Models and methods}

\subsection{Structures}

We use a 216-atom diamond-structure crystal, cSi-216, as the reference, and amorphous networks that differ in size, preparation, and coordination-defect content (Table~\ref{tab:si_models}). DTW-4096 and BM-10K were relaxed with the Si/H Atomic Cluster Expansion (ACE) potential of Rosset and Deringer~\cite{Rosset2026ACE}, which is based on the atomic cluster expansion~\cite{Drautz2019ACE,Lysogorskiy2021PACE} and was run in LAMMPS~\cite{Thompson2022LAMMPS}. The other amorphous structures are used as supplied: DTW-4096-U, DTW-512, the electronically defective aSi-216 and aSi-512 cells, and R10 and R2--R4. A short ACE minimization shows that R2--R4 are already close to ACE minima, while the others are not (Sec.~S1). TB runs on ACE-relaxed versions of the others leave the aggregate statistics unchanged but change the atom-resolved maps and which runs show a significant filament (Table~S12). The comparison with the thermal study uses the ACE-relaxed DTW-4096 and ACE-relaxed versions of R2--R4.

The large networks are the 512- and 4096-atom Djordjevi\'c--Thorpe--Wooten models, DTW-512 and DTW-4096~\cite{Djordjevic1995DTW}, and a 10\,000-atom Barkema--Mousseau network, BM-10K~\cite{Barkema2000highquality}. We use both the original DTW-4096 configuration (DTW-4096-U) and its ACE-relaxed counterpart, DTW-4096. The four 216-atom models (aSi-216-A--D) and three 512-atom models (aSi-512-A--C) carry coordination defects. An atom is coordination-defective when its number of neighbours within 2.8~\AA\ is not four. The bond clusters of Sec.~IV~G use instead each cell's first minimum of $g(r)$ to define first-neighbour bonds (Sec.~S3); for DTW-4096-U this is 2.57~\AA, at which it is entirely fourfold.

R10 and R2--R4 come from the machine-learning molecular-dynamics dataset of Rosset \textit{et al.}~\cite{Rosset2025paracrystallinity} (see also Ref.~\cite{Treacy2012}), generated by melt--quench runs at different cooling rates and densities. R10 has 216 atoms and is entirely fourfold coordinated, with a clean electronic gap and structure consistent with experiment~\cite{Rosset2026defectfree}. R2--R4 are paracrystalline cells of 1000 atoms with coordination-defect fractions of 1.0\%, 0.8\%, and 0.6\%; R2 has only threefold defects, and R3 and R4 have threefold and fivefold sites. Together these models let us compare fourfold networks with under- and over-coordinated ones.

\subsection{Hamiltonians and algorithm}

The tight-binding (TB) model is the orthogonal $sp^3$ model of Kwon \textit{et al.}~\cite{Kwon1994TB}, a Slater--Koster-type parametrization~\cite{Slater1954,Harrison1980} used with its published hopping scalings and cutoffs. Our implementation reproduces the reported crystalline-Si gaps: the indirect gap of 0.78~eV and the direct gap of 1.62~eV at $\Gamma$. The 216-atom cell at $\Gamma$ samples only a $3\times3\times3$ mesh of the cubic Brillouin zone, so its gap is larger (1.17~eV). The DFT Hamiltonians come from SIESTA v5.4.2~\cite{Jose2002SIESTA_1,Garcia2020SIESTA_2}, with the local-density approximation (Ceperley--Alder, Perdew--Zunger)~\cite{Ceperley1980,Perdew1981}, a double-$\zeta$ basis, and $\Gamma$-point sampling, followed by L\"owdin orthogonalization~\cite{Aiken1980Lowdin}. The velocity operator is built from atom-centre minimum-image separations, the usual tight-binding prescription~\cite{Graf1995}. Unless stated otherwise, $T=300~\mathrm{K}$ and $\eta=0.1~\mathrm{eV}$.

The two Hamiltonians differ. The Kwon model is two-centre, non-self-consistent, and parametrized for $sp^3$ silicon. The SIESTA Hamiltonians are self-consistent, use a non-orthogonal localized basis, and include longer-range hopping. We therefore expect broad spatial patterns to transfer between them better than features tied to single defect levels or small bond-level differences. We do not assume agreement with plane-wave calculations such as Ref.~\cite{Ugwumadu2026KuboMap}. Basis-set convergence was tested once (Sec.~S1): replacing the double-$\zeta$ Si basis by single-$\zeta$ in one hydrogenated cell (1000 atoms, 10 at.\% H) leaves the per-atom maps highly correlated (Spearman 0.88--0.99 within 0.1~eV of the band edges) but changes $\sigma$ by 6--10\%.

Table~S1 of the Supplemental Material lists the 136 windowed TB runs and 55 DFT runs. A run is one calculation for a given structure, Hamiltonian, chemical potential, and broadening; the several maps obtained from it are not additional runs.

To reduce the cost, the windowed TB calculations keep only the eigenstates that satisfy
\begin{equation}
|\epsilon_n-\mu|
\leq 1.5\left(15k_BT+5\eta\right),
\end{equation}
They are found by ARPACK shift-and-invert iteration with a sparse LU factorization~\cite{Lehoucq1998ARPACK}, and contributions that need states outside the window are approximated by a Neumann-series expansion. The outermost 15\% of the window contributes at most $4.6\times10^{-10}$ of $\sigma$ over the 136 TB runs (at most $2.1\times10^{-9}$ in the 29 later runs; Table~S1, Sec.~S1). This diagnostic measures the weight near the edge of the retained window; it is not a bound on the truncation error. The windowed method handles up to 10\,000 atoms in 5--7~GB of memory and is described further in Ref.~\cite{Drabold2026sdos}. Order-$N$ methods~\cite{Goedecker1999,Stephan1998OrderN}, including the solver in SIESTA~\cite{Jose2002SIESTA_1}, avoid diagonalization; we use a windowed eigenvalue method because the analysis needs individual eigenstates. The DFT calculations use dense diagonalization.

Against dense calculations, the windowed and dense maps differ by at most $3\times10^{-12}$ in the cases tested. Finite-difference derivatives agree to $\leq8\times10^{-8}$, the sum rules to $\leq3\times10^{-8}$ ($\leq5\times10^{-5}$ for DFT quantities, whose derivatives are evaluated by finite differences), and the $\sigma$-force to $6\times10^{-7}$ relative to finite differences.

\begin{figure*}[!t]
\includegraphics[width=\linewidth]{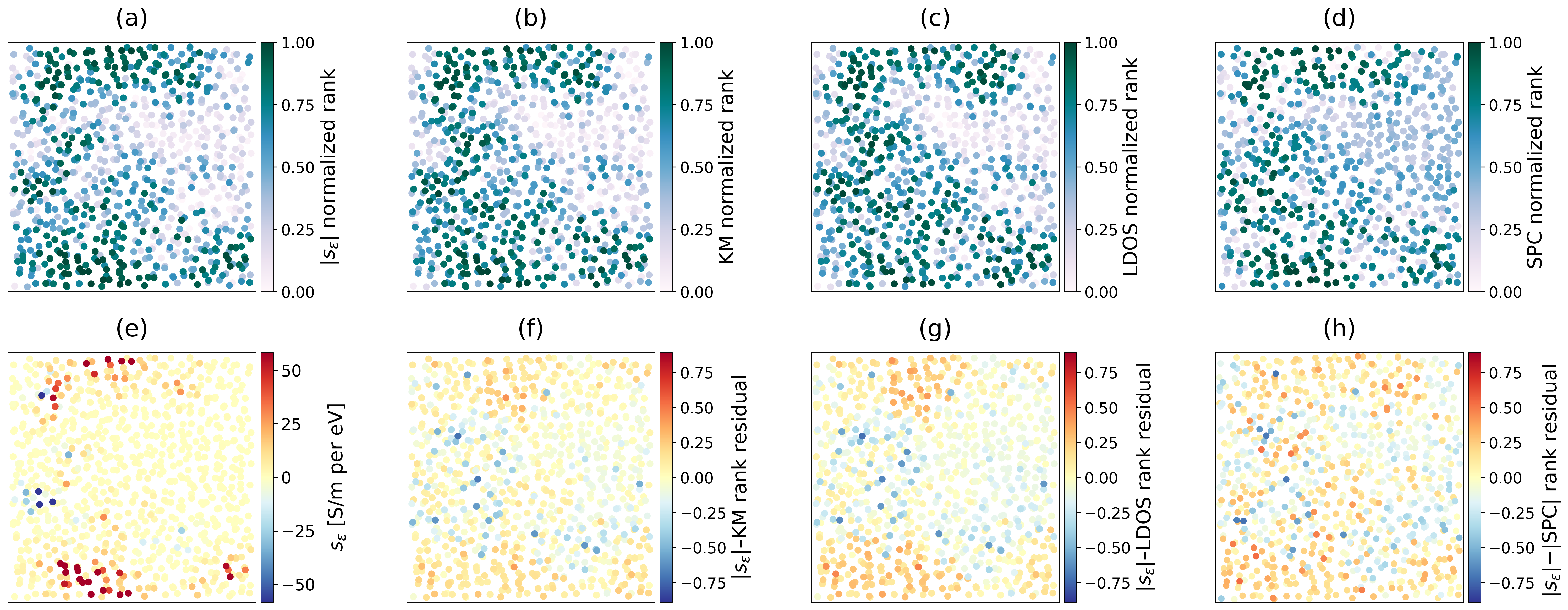}
\caption{Comparison of conductivity maps within the 863-atom slab $15<z<24$ \AA{} of ACE-relaxed DTW-4096 at $\mu=0.2~\mathrm{eV}$. (a--d) Normalized ranks of sensitivity magnitude $|s_\varepsilon|$, KuboMap, LDOS, and signed SPC, respectively. (e) Signed on-site sensitivity, with symmetric color limits set by the full-cell 98th percentile of $|s_\varepsilon|$. (f--h) Sensitivity-magnitude rank residuals after separate linear regressions against KuboMap, LDOS, and $|\mathrm{SPC}|$ ranks, respectively (panel (d) shows the signed SPC). Ranks and regression coefficients are calculated over the full supercell.}
\label{fig:maps}
\end{figure*}

\subsection{Map comparison}

We compare maps by rank agreement, by overlap of the top-ranked atoms, and by recovery of the conductivity sensitivity. Every comparison is made on the same configuration, at the same chemical potential and broadening.

Let $m_1$ and $m_2$ be two maps defined on the same $N$ atoms, with value $m_{\ell,i}$ at atom $i$ ($\ell=1,2$). The normalized atomic ranks are
\begin{equation}\label{eq:rank}
R_i(m_\ell)
=
\frac{\operatorname{rank}(m_{\ell,i})}{N},
\end{equation}
with average ranks for ties. Spearman's correlation is then~\cite{Wissler1905Spearman}
\begin{equation}
\rho_s(m_1,m_2)
=
\operatorname{Corr}\!\left(
\{R_i(m_1)\}_{i=1}^{N},
\{R_i(m_2)\}_{i=1}^{N}
\right).
\label{eq:spearman}
\end{equation}
It compares orderings and ignores the scale of each map. We evaluate it at single sites and after averaging each map over 4~\AA\ neighbourhoods and ranking the averages.

Agreement between top-ranked sets is measured by the overlap fraction
\begin{equation}
O_k(m_1,m_2)
=
\frac{
|\mathcal T_k(m_1)\cap\mathcal T_k(m_2)|
}{k},
\label{eq:top-set-overlap}
\end{equation}
where $\mathcal T_k(m)$ is the set of indices of the $k$ highest-ranked atoms in map $m$. Both sets contain exactly $k$ atoms, so $O_k$ runs from zero for disjoint sets to one for identical sets.

To measure how well a map $m$ finds the sites of strong on-site sensitivity, we define its sensitivity recovery
\begin{equation}
\mathcal R_k(m)
=
\frac{
\displaystyle\sum_{i\in\mathcal T_k(m)}
|s_{\varepsilon,i}|
}{
\displaystyle\sum_{i\in\mathcal T_k(|s_\varepsilon|)}
|s_{\varepsilon,i}|
}.
\label{eq:sensitivity_recovery}
\end{equation}
Here $m$ may be LDOS, KuboMap, or SPC. The denominator is the largest total sensitivity magnitude that $k$ atoms can capture, so $0\leq\mathcal R_k(m)\leq1$ whenever it is nonzero, and unity means optimal recovery. We take $k$ to be the top 1\% and 5\% of atoms and compare with random selections of the same size. Unlike the Spearman correlation and the overlap $O_k$, this metric uses $|s_\varepsilon|$ as its reference and weights the selected atoms by their sensitivity magnitudes.

We estimate variability with a spatial block bootstrap~\cite{Kunsch1989,Lahiri2003} of 64 cubic blocks (4$\times$4$\times$4) and 400 resamples (300 for the correlation intervals of the 30-map set). Blocks are drawn with replacement and the metrics are recomputed for each realization. Both maps in a comparison use the same resampled blocks, which keeps the spatial dependence inside each block. Comparisons that involve sensitivity use $|\se|$ unless stated otherwise, so they test the magnitude of the response and not whether an on-site perturbation raises or lowers $\sigma$.

\section{Results and Discussion}

Sections IV~A and IV~G--IV~H carry the two main results; Secs.~IV~B--IV~F audit the allocation maps against the first, and Sec.~IV~I compares with the thermal maps.

\subsection{The sensitivity is the exact response}\label{subsec:single_atom_perturb}

We first test the sensitivity against brute force. We recomputed $\sigma$ exactly, by dense diagonalization, for every atom of two defective aSi-216 supercells (aSi-216-A and aSi-216-B, eight values of $\mu$ each) and of DTW-512 (five values), either removing the atom or shifting its on-site energy by $+0.3$~eV. For the shift, $|\se|$ ranks the atoms by their actual response at Spearman 0.97--1.00, with top-10\% overlap 0.78--1.00. The corresponding ranges are 0.33--0.90 for LDOS, 0.47--0.81 for KuboMap, and 0.31--0.71 for SPC. With LDOS regressed out, the partial correlation stays at 0.97--1.00 for $|\se|$ and falls to $-0.30$--0.72 for KuboMap and $-0.01$--0.38 for SPC. The exact response is close to the linear prediction $s_\varepsilon\delta$ for most atoms (median ratio 0.90--1.08 over atoms; at most 3.0\% of the atoms have the wrong sign), but not for the atoms that respond most: weighted by $|\se|$ the median ratio is 0.79--1.53 and the upper quartile reaches 3.3.

Atom removal is different, and the outcome depends on $\mu$ and on the cell. In the defective aSi-216 models no map predicts the response: correlations of $|\Delta\sigma/\sigma|$ with the maps lie between $-0.39$ and $+0.39$, and the top-10\% overlap is 0.00--0.24 against 0.10 for random choice, except at $\mu=-0.129$~eV in the second cell (0.29--0.38). In DTW-512 some energies are predicted and others are not: $|\se|$ correlates 0.76 with $|\Delta\sigma/\sigma|$ at $\mu=1.1$~eV and 0.48 at $-0.1$~eV, and not at all at 0.1 and 0.3~eV ($|\rho|\le0.03$). Removing an atom creates dangling-bond states, and none of the maps describes them.

\subsection{Maps and the sign of the response}

Having fixed the reference, we ask how well the allocation maps reproduce it. Figure~\ref{fig:maps} compares the local density of states (LDOS), the KuboMap and SPC partitions, and the on-site conductivity sensitivity in one 9~\AA-thick slab of ACE-relaxed DTW-4096. Ranks and fits use all $N=4096$ atoms. Panels (a--d) show normalized ranks [Eq.~\eqref{eq:rank}], so maps with different units can be compared. Panel (e) keeps the signed sensitivity $s_\varepsilon$, which is positive when $\sigma$ rises as the on-site energy of that atom is raised.

KuboMap and LDOS are closely aligned: their full-cell Spearman correlation is 0.963. The sensitivity magnitude correlates with KuboMap at 0.815 and with LDOS at 0.779. The maps agree substantially on where electronic weight and response sit, but they are not interchangeable. SPC differs more: its correlations with LDOS, KuboMap, and sensitivity magnitude are 0.385, 0.393, and 0.373.

SPC is also signed. Here 33.8\% of the atoms have negative SPC, and they carry 13.0\% of the total absolute SPC weight. The sign of SPC is the sign of a contribution to the partition; the sign of $s_\varepsilon$ is the sign of a response to a perturbation. They are different things. Panel (d) ranks the signed values, so a low rank can mean a strongly negative contribution.

Panels (f--h) show where the sensitivity ordering departs from what each map predicts. For each predictor $m \in \{\rho_{\mathrm{km}},\rho_{\mathrm{ldos}},\sigma_{\mathrm{spc}}\}$ we fit
\begin{equation}
\begin{aligned}
(\widehat a_m,\widehat b_m)
&=
\underset{a,b}{\operatorname{arg\,min}}
\sum_{i=1}^{N}
\left[
R_i(|s_\varepsilon|)-aR_i(m)-b
\right]^2,\\
\Delta r_i^{(m)}
&=
R_i(|s_\varepsilon|)
-\left[\widehat a_m R_i(m)+\widehat b_m\right].
\end{aligned}
\label{eq:sensitivity-rank-residual}
\end{equation}
$\Delta r_i^{(m)}$ is positive when the sensitivity-magnitude rank of atom $i$ is higher than map $m$ predicts and negative when it is lower. A residual near zero means the atom follows the fit, not that it is insensitive. Because the regression uses $|s_\varepsilon|$, the sign of a residual says nothing about whether $\sigma$ rises or falls. In panel (h) the predictor is $|\sigma_{\mathrm{spc}}|$, which tracks the sensitivity magnitude more closely than the signed SPC of panel (d).

In this configuration the in-sample $R^2$ is 0.664 for KuboMap, 0.608 for LDOS, and 0.413 for $|\mathrm{SPC}|$ (0.139 for the signed SPC). These are descriptive fits within one configuration, not cross-validated predictions. With the LDOS rank removed from both the sensitivity magnitude and each predictor, the partial rank correlations are 0.379 for KuboMap and 0.258 for $|\mathrm{SPC}|$ (0.126 for the signed SPC). KuboMap carries information about the sensitivity ordering beyond what it shares with the LDOS; $|\mathrm{SPC}|$ adds less, and the signed SPC least.

We repeated the comparison for 84 maps of models with at least 1000 atoms (TB; $\eta=0.1$~eV; $\mu=-2$ to $+3.5~\mathrm{eV}$; the original and ACE-relaxed 4096-atom networks, the 10\,000-atom network, and the paracrystalline 1000-atom cells in their original form and on the ACE-relaxed structures of the thermal study~\cite{Drabold2026SPTC}). The KuboMap--LDOS Spearman correlation has median 0.97 (range 0.56--1.00). For $|\se|$ the median correlation is 0.80 (0.05--0.89) with LDOS and 0.81 ($-0.23$--0.91) with KuboMap. Here SPC enters as its magnitude: $|\mathrm{SPC}|$ correlates with LDOS at 0.67 (0.29--0.80), and the signed SPC at 0.41 (0.10--0.80). The agreement is not universal. In paracrystalline R2 the $|\se|$--LDOS correlation is only 0.07--0.43, and in the gap and deep inside the bands it falls to 0.4 or less. With the LDOS rank removed as in Eq.~\eqref{eq:sensitivity-rank-residual}, the residual correlation of $|\se|$ has median 0.27 with KuboMap and 0.23 with $|\mathrm{SPC}|$ (0.14 with signed SPC).

Can simple local structure predict the maps? We regressed the maps linearly on five local structural descriptors of each atom (Sec.~S8), for 30 maps spanning $0\leq\mu\leq0.5~\mathrm{eV}$, with five-fold spatial-slab cross-validation. From an atom's own descriptors the cross-validated $R^2$ is 0.06 for the LDOS rank, 0.07 for KuboMap, and 0.03 for the sensitivity-magnitude rank. Adding descriptor averages over neighbourhoods of radius 4--6~\AA{} raises these to 0.16--0.18, 0.17--0.18, and 0.08--0.09. Structure captures only a small part of the variation, least of all for the sensitivity, and these descriptors cannot replace the electronic calculation.

Magnitude and direction of the response are separate matters. At the chemical potential of Fig.~\ref{fig:maps}(e), 81.6\% of the sites have positive $s_\varepsilon$, and the rest respond with the opposite sign, even within regions of high sensitivity magnitude. Over the valence-tail cases we analysed, 78--100\% of sites are positive; in the conduction-tail cases 91--100\% are negative; near defect levels the sign can alternate. The sign is not fixed by $\mu$ alone, and agreement between the magnitudes of two maps does not mean they agree on the direction of the response.

\begin{figure*}[!t]
\includegraphics[width=\textwidth]{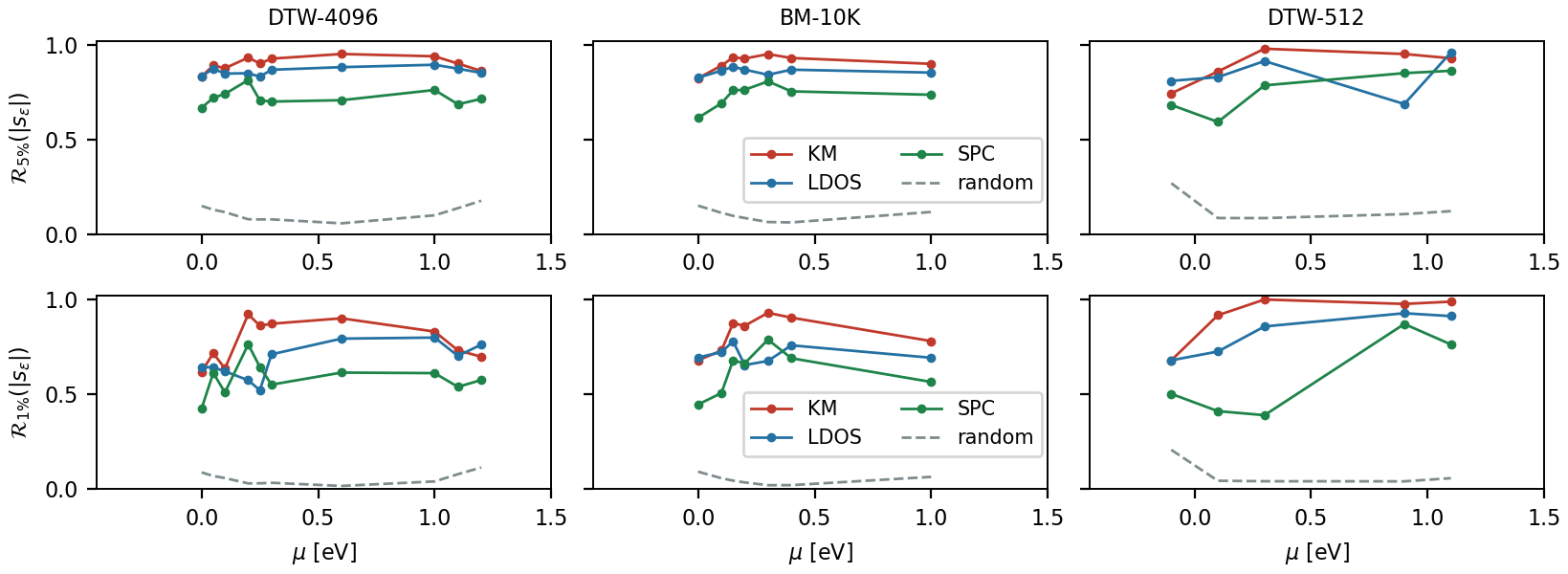}
\caption{Sensitivity recovery of the top 5\% (top) and top 1\% (bottom) of $|\se|$ by top-ranked KuboMap, LDOS, and SPC atoms, against $\mu$, in three tight-binding supercells; dashed lines show random selection. }
\label{fig:capture}
\end{figure*}

\subsection{How well do partitions recover where \texorpdfstring{$\sigma$}{sigma} responds?}

Figure~\ref{fig:capture} gives the sensitivity recovery $\mathcal R_k(m)$ [Eq.~\eqref{eq:sensitivity_recovery}] against $\mu$: the share of the largest attainable sensitivity magnitude that the top-ranked atoms of map $m$ capture. In the valence tail the top 5\% of KuboMap atoms collect 0.88--0.95 of the top-5\% sensitivity (random: 0.05--0.10); LDOS collects 0.84--0.91 and SPC 0.70--0.81. The bootstrap intervals exclude zero for most differences. KuboMap minus LDOS is $+0.03$ to $+0.11$ at top 5\% (for example $+0.08$ [$0.05,0.13$] at $\mu=0.2$~eV in relaxed 4096), and rises to $+0.35$ at top 1\% in a narrow window above the valence edge; KuboMap minus SPC is $+0.12$ to $+0.23$. Far inside the bands, and in the clean-gap R10, LDOS does as well as KuboMap.

Collective shifts are a different matter. Shifting the on-site energy of the 20 top-ranked DTW-512 atoms by $+0.1$~eV changes $\sigma$ by the fractions $\Delta\sigma/\sigma_0=+7.16$ ($\mu=0.3$~eV), $-0.75$ (0.9~eV), and $-0.38$ (1.1~eV) for the KuboMap-ranked atoms. $\Delta\sigma/\sigma_0$ is dimensionless: $+7.16$ means that $\sigma$ rises by a factor of 8.2. The $|\se|$ ranking gives $+7.22$, $-0.76$, and $-0.39$; a random selection of 20 atoms changes $\sigma$ by 0.22, 0.15, and 0.07 on average in $|\Delta\sigma/\sigma_0|$ (40 draws), so the ranked sets exceed it by factors of 5--32. At $\mu=0.1$~eV the KuboMap-ranked atoms carry both signs and change $\sigma$ by almost nothing ($-0.02$): KuboMap gives the magnitude of the response, not its direction. The response of these sets is not linear: at $\mu=0.3$~eV the linear fraction of the $|\se|$ set is 0.90, 0.71, 0.28, and 0.023 for $\delta=0.01$, 0.03, 0.1, and 0.3~eV. Twenty atoms shifted together move the conductivity of a tail state far beyond first order (Sec.~S8).

\subsection{SIESTA Hamiltonians}
\begin{figure}[!t]
\includegraphics[width=\linewidth]{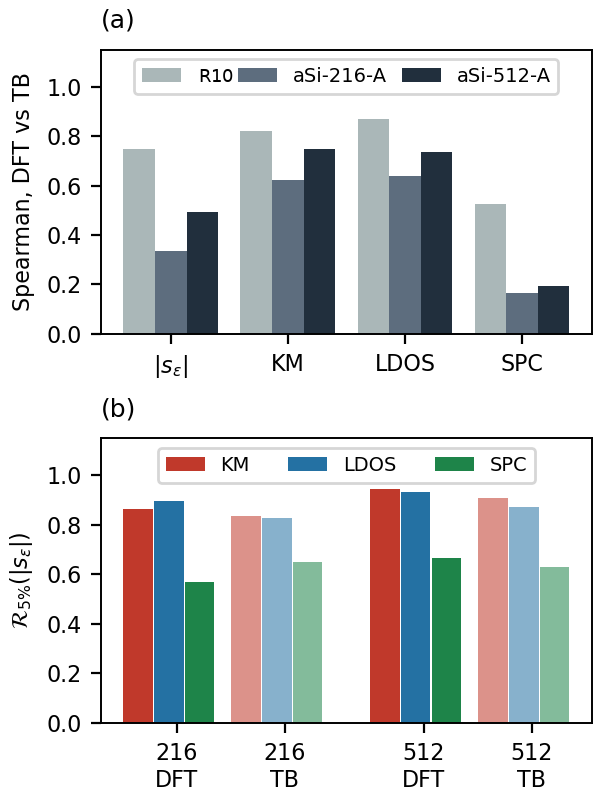}
\caption{(a) Median Spearman correlation between DFT and tight-binding maps at matched energies. (b) Median top-5\% sensitivity recovery of $|\se|$ by KuboMap, LDOS, and SPC in DFT (solid) and tight-binding (faded) models. }
\label{fig:dft}
\end{figure}

We repeated the analysis with SIESTA Hamiltonians (LDA, double-$\zeta$) for R10 ($\Gamma$ gap 1.41~eV), four defective aSi-216 supercells, and three defective aSi-512 supercells (gaps 0.012--0.082~eV), matching energies by distance from each cell's band edge. In R10, DFT and TB maps agree at Spearman 0.75--0.87 at the valence edge and 0.41--0.52 at the conduction edge, for $\se$, KuboMap, and LDOS [Fig.~\ref{fig:dft}]. In the defective models the smooth maps transfer with median Spearman 0.62 and 0.75 for KuboMap and 0.64 and 0.74 for LDOS (216 and 512 atoms). Atom-level sensitivity transfers moderately (0.34 and 0.49) and SPC poorly (0.17 and 0.19). The defect levels fall at different energies in the two Hamiltonians.

In DFT the median $\mathcal R_{5\%}(m)$ is 0.86 for KuboMap, 0.90 for LDOS, and 0.57 for SPC with 216 atoms, and 0.95, 0.93, and 0.67 with 512 atoms. KuboMap beats SPC in 26/29 and 20/20 runs, and LDOS beats KuboMap in 11/29 and 13/20. In the DFT supercells KuboMap therefore has no consistent advantage over LDOS.

\subsection{Gauge ambiguity of KuboMap and the binned alternative}
\label{subsec:gauge_binned_KM}

KuboMap built from individual eigenstates depends on the choice of eigenvectors in degenerate subspaces. We test this with random rotations of the eigenvectors. Within exactly degenerate multiplets these leave the Hamiltonian and its eigenbasis valid. Where exact degeneracies are absent, we rotate within near-degenerate pairs and within energy bins of width $\eta/10$ and $\eta/3$ (Haar-random~\cite{Mezzadri2007Haar}); these rotations are not gauge transformations but measure how much the map depends on the basis. We call the map of Eq.~\eqref{eq:partitions_km} the state-resolved KuboMap and that of Eq.~\eqref{eq:binned_KM} the spectrally binned KuboMap. Tables~S7 and~S8 give the details.

In crystalline cSi-216 at $\Gamma$ every atom is equivalent. LDOS and SPC are uniform to $3\times10^{-7}$ (DFT) and $2\times10^{-14}$ (TB) in any gauge. The state-resolved KuboMap is not uniform: its atom-to-atom coefficient of variation is 1--4\% (DFT) and 2--14\% (TB), and this variation is purely gauge-induced. In the amorphous cells the dependence ranges from negligible (relative $L_2\leq6\times10^{-4}$ within 0.3~eV of the aSi-216 band edges) to 0.07--0.29 elsewhere. Averaging over spheres of radius up to 12~\AA{} still leaves relative changes of 0.05--0.07, so the dependence is an extended change of the map, not atom-scale noise (Sec.~S8).

The spectrally binned KuboMap (Appendix~\ref{app:binned}) is uniform in the crystal to $10^{-8}$ and invariant under rotations to $9\times10^{-16}$. It costs two things: closure is approximate ($\sigma$ is reproduced to 0.04\%, about 1\%, and up to 7\% for $\delta=\eta/30$, $\eta/10$, and $\eta/3$), and the map depends on where the bin edges fall (shifts change it by 0.04--0.07 at $\delta=\eta/10$ and up to 0.52 at $\eta/3$). Binning moves the ambiguity from the eigenvectors to the spectral partition. It does not make the map more stable against physical perturbations: random on-site shifts of 10--30~meV change all versions by comparable amounts (relative $L_2$ 0.06--0.15).

SPC makes the opposite trade: it is gauge invariant but signed. In 30 TB maps of cells with at least 1000 atoms, a median 34\% of atoms (maximum 44\%) have SPC of opposite sign to $\sigma$, and they carry 16\% of $\sum|\mathrm{SPC}|$. Averaging over 8~\AA{} reduces this fraction to 2\% (maximum 25\%).

All of these ambiguities are small compared with thermal disorder. In 20 snapshots of DTW-4096 at 300~K, $\sigma$ is 2.2--2.9 times its relaxed value, with relative standard deviation 16--24\%. The top-1\% sets of all four maps overlap the relaxed ones by only 0.22--0.37 (Table~S4). The KuboMap gauge changes the map by 0.05--0.07.

\subsection{Position sensitivity}

Position derivatives give the response of $\sigma$ to an infinitesimal displacement of an atom. In DTW-4096 at $\mu=0.2~\mathrm{eV}$ the magnitude $\|\partial\sigma/\partial\mathbf X_x\|$ correlates with the on-site sensitivity magnitude $|\se(x)|$ at Spearman 0.85 (Fig.~S3a): the atomic rankings largely agree, although the two derivatives describe different perturbations. The bond-position sensitivity splits into a radial part (stretching) and a transverse part (infinitesimal rotation); a ratio $s_\perp/|s_\parallel|$ below one means a larger radial response (Fig.~S3b). The ratio of RMS component magnitudes, $Q_{\perp/\parallel}=(\sum_bs_{\perp,b}^2/\sum_bs_{\parallel,b}^2)^{1/2}$, is 1.55 at $\mu=0.2~\mathrm{eV}$ and 0.59 at $\mu=1.0~\mathrm{eV}$: the overall response is mostly transverse at the first energy and mostly radial at the second. Rotating a bond carries information that stretching alone does not.

\subsection{Topology of sensitive bond networks}

Are the most sensitive bonds scattered, or do they connect? At each chemical potential $\mu$ we select the 1\% of bonds with the largest sensitivity magnitudes $|\sbond|$. Two selected bonds are connected when they share an atom, as in cluster statistics for percolation~\cite{Stauffer1994}. We measure the clustering by $L_{\mathrm{obs}}$, the number of bonds in the largest connected cluster.

Is this clustering just the concentration of electronic weight? To find out, we generate random bond selections of the same size, weighted by the LDOS. $L_{\mathrm{LDOS}}$ is the largest-cluster size in one such selection, and $\langle L_{\mathrm{LDOS}}\rangle$ is its mean over repeated selections. We define the standardized cluster excess as
\begin{equation}
z_{\mathrm{LDOS}}
=
\frac{
L_{\mathrm{obs}}-\langle L_{\mathrm{LDOS}}\rangle
}{
\operatorname{std}(L_{\mathrm{LDOS}})
},
\label{eq:cluster-excess}
\end{equation}
where $\operatorname{std}$ is the standard deviation over the same selections (the score is defined only when it is nonzero). A positive score means clusters larger than the reference mean, in units of the reference standard deviation. $z_{\mathrm{LDOS}}$ measures excess clustering, not sensitivity magnitude.

\begin{figure}[!t]
\includegraphics[width=\columnwidth]{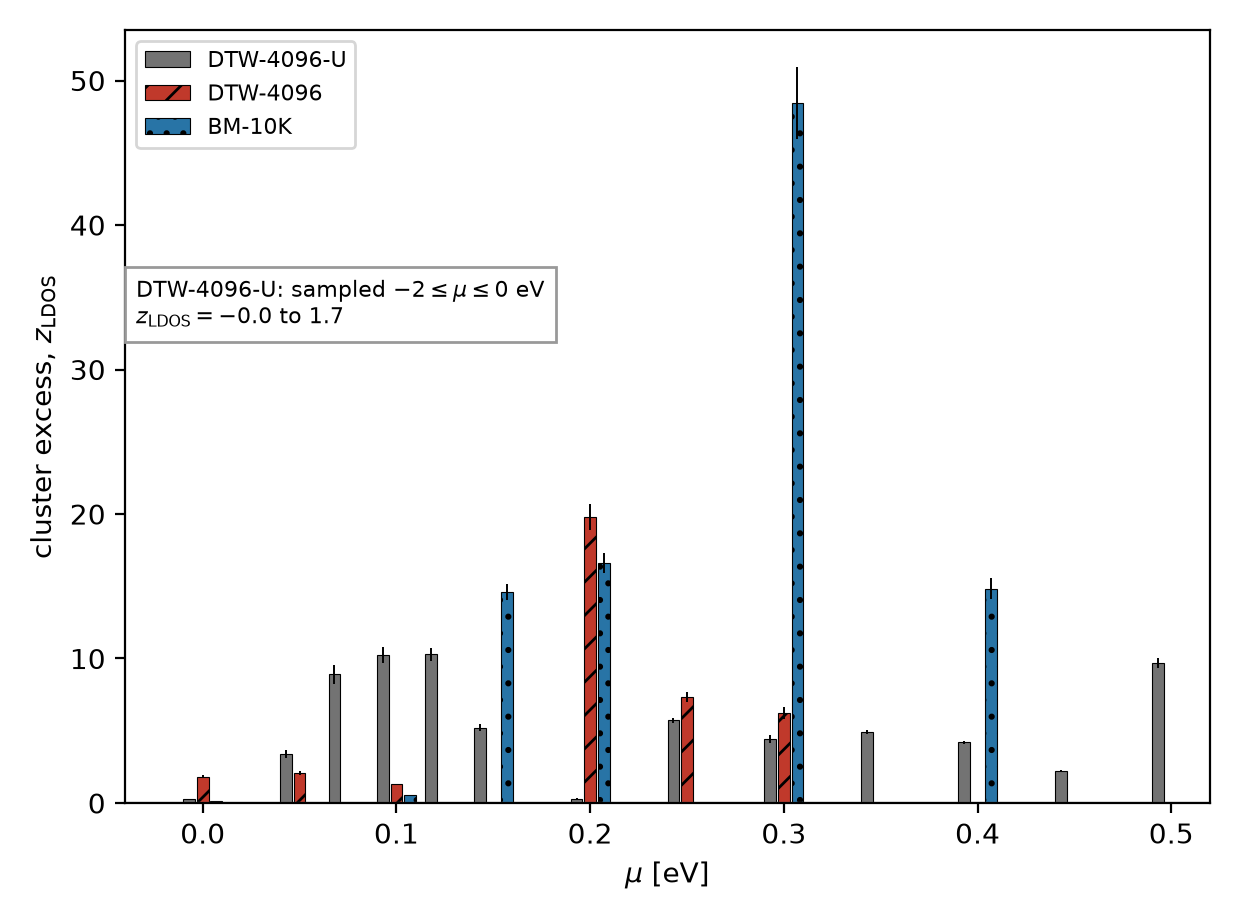}
\caption{Standardized largest-cluster excess $z_{\mathrm{LDOS}}$ [Eq.~\eqref{eq:cluster-excess}] for the top-1\% $|\sbond|$ bonds as a function of chemical potential. The reference consists of LDOS-weighted random bond selections containing the same number
of bonds. Selected energies include the maximum score for each model. Bars show the mean over ten sets of 400 selections and error bars the s.d.\ over the sets. Bars are offset horizontally for clarity; missing bars indicate unavailable samples at the selected energies.}
\label{fig:clusters}
\end{figure}

Figure~\ref{fig:clusters} shows that the excess depends strongly on $\mu$ and on the structure (each $z_{\mathrm{LDOS}}$ is a mean $\pm$ s.d.\ over ten sets of 400 reference selections). DTW-4096-U reaches $10.3\pm0.4$ at $\mu=0.125~\mathrm{eV}$, DTW-4096 reaches $19.8\pm0.9$ at $0.200~\mathrm{eV}$, and BM-10K reaches $48.5\pm2.5$ at $0.300~\mathrm{eV}$. The largest clusters contain 23 bonds against a reference mean of 6.1 in DTW-4096-U, 37 against 7.5 in DTW-4096, and 39--114 against 9--13 over $0.20$--$0.30~\mathrm{eV}$ in BM-10K. Further excess appears near $0.50~\mathrm{eV}$ in DTW-4096-U ($9.7\pm0.3$) and near $1.0~\mathrm{eV}$ in DTW-4096 ($6.2\pm0.3$, not in the plot). The largest cluster is the component with the most bonds throughout (Sec.~S3).

Over 120 TB maps with $-0.2\leq\mu\leq0.8~\mathrm{eV}$, 26 have $z_{\mathrm{LDOS}}>3$. Excess clustering occurs in all three large networks and weakly in one aSi-512 cell. A few smaller-cell maps also exceed three reference standard deviations, but their selected sets are too small to show comparable extended clustering. Size matters here. The top 1\% holds only 4, 10, and 19--20 bonds in the 216-, 512-, and 1000-atom supercells, against about 82 and 200 in the 4096- and 10\,000-atom structures, and small selections limit the clusters and connectivity one can resolve. The 216- and 512-atom cells and DTW-4096-U are used as supplied, not at ACE minima. On ACE-relaxed versions of the aSi cells the significant runs are not the same ones (8 of 50 original runs in this window, 6 of 50 relaxed, none in both; Table~S12), so the counts for these cells describe the structures as supplied. The ACE-relaxed DTW-4096 also shows a pronounced excess ($z_A$ up to 19.8), at different $\mu$ (Table~II).

How robust is the excess? Ten independent sets of 400 reference selections change the scores by a standard deviation of 0.03--2.5 in the nine cases of Tables~S9 and~S10, and the prominent peaks remain. Changing the selected bond fraction from 1\% to 0.5\% or 2\% can change a score by up to a factor of 2, so its magnitude depends on the threshold.

The excess also depends on the broadening and on temperature. In DTW-4096 at 0.2~eV, $z_{\mathrm{LDOS}}$ is $0.5\pm0.1$, $19.8\pm0.9$, and $26.7\pm1.1$ for $\eta=0.05$, 0.10, and 0.20~eV. At $\eta=0.05$~eV the largest cluster has 14 bonds against a reference mean of 13.0: the filament is not distinguishable from the reference. The other cases follow the same trend (Table~S11). In every case with a dense spectrum near $\mu$, $z_{\mathrm{LDOS}}$ exceeds 3 for $\eta/\Delta\geq5.2$ and is below 3 for $\eta/\Delta\leq3.5$, where $\Delta$ is the mean level spacing at $\mu$. (At 0.5~eV in DTW-4096-U only two states lie within 0.1~eV of $\mu$, and $\Delta$ is not defined by a dense spectrum.) In the 20 thermal snapshots at 0.2~eV the largest cluster has $11\pm5$ bonds (37 in the relaxed cell), $z_{\mathrm{LDOS}}=4.2\pm3.3$ (relaxed: $19.8\pm0.9$), and contains $18\pm19$\% of the atoms of the relaxed cluster. A random displacement of 0.05~\AA{} of every atom keeps part of it: 20 bonds, $z_{\mathrm{LDOS}}=9.5\pm0.6$, and 57\% of the atoms. At 0.1~eV, where the relaxed network shows no excess ($z_{\mathrm{LDOS}}=1.3\pm0.0$), the snapshots show none ($1.7\pm1.0$). The filaments are a property of the relaxed networks at the broadening we used.

Nor is the clustering special to the sensitivity. Bond rankings from KuboMap or SPC give comparable or larger clusters in several energy windows. At $\mu=0.20~\mathrm{eV}$ in BM-10K the KuboMap, SPC, and $\sbond$ selections have largest clusters of 22, 41, and 39 bonds, and near the three principal peaks $\sbond$ gives larger clusters than the other maps. The excess therefore reflects a spatial organization shared by several electronic response measures, with stronger sensitivity-based clustering in particular energy windows.

\begin{figure*}[!t]
\includegraphics[width=0.72\textwidth]{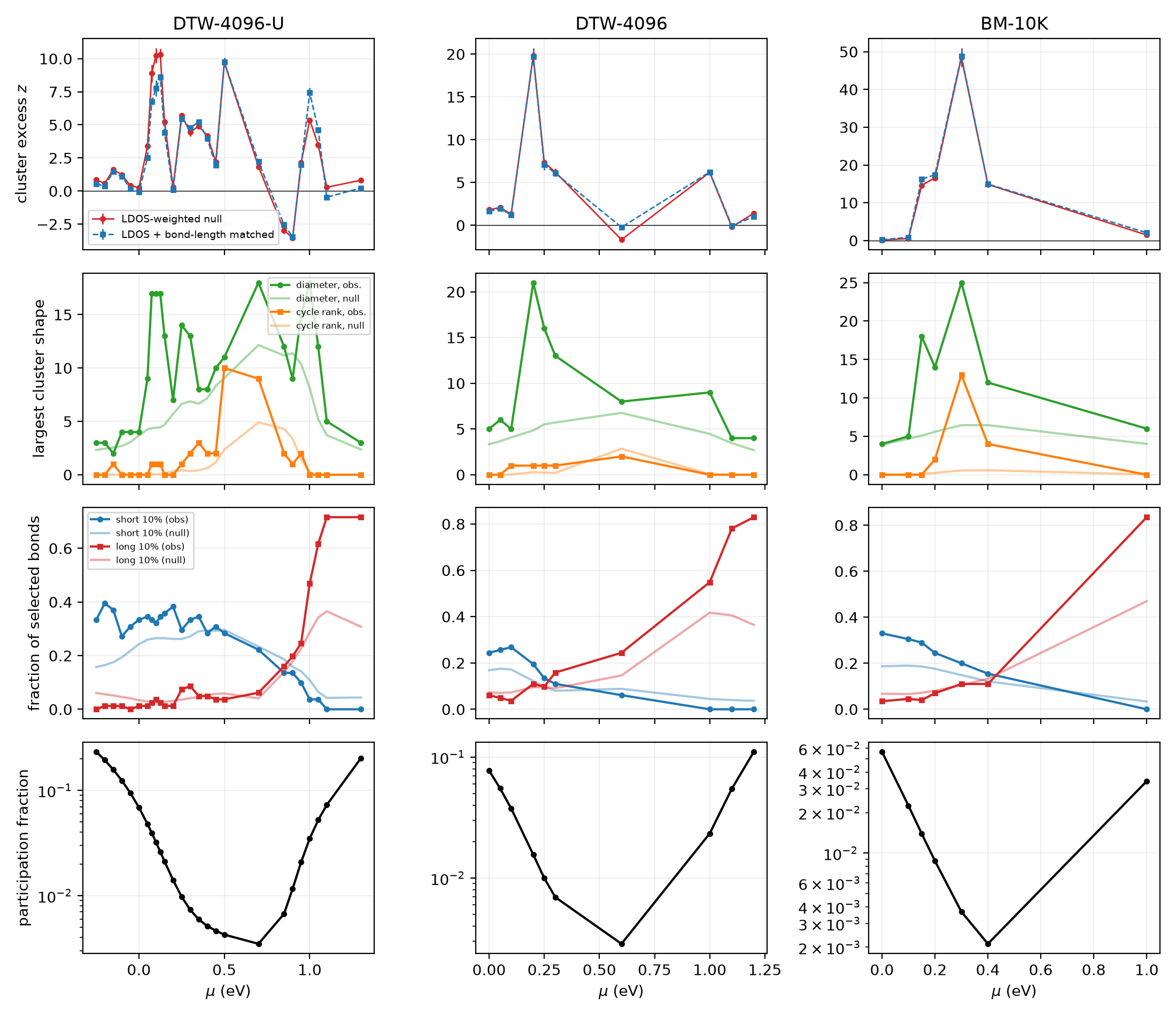}
\caption{Cluster topology against chemical potential for the top-1\% $|\sbond|$ bonds in DTW-4096-U, DTW-4096, and BM-10K (400 reference selections per point). Row 1: cluster excess $z$ against the LDOS-weighted reference (red) and against a reference that also matches the bond-length distribution (blue). Row 2: diameter and cycle rank of the largest cluster (dark) and the reference means (light). Row 3: fraction of selected bonds among the shortest (blue) and longest (red) 10\% of bonds, with reference means (light). Row 4: participation fraction of the states within $\eta$ of $\mu$.}
\label{fig:topology}
\end{figure*}

\begin{figure*}[!t]
\includegraphics[width=0.8\textwidth]{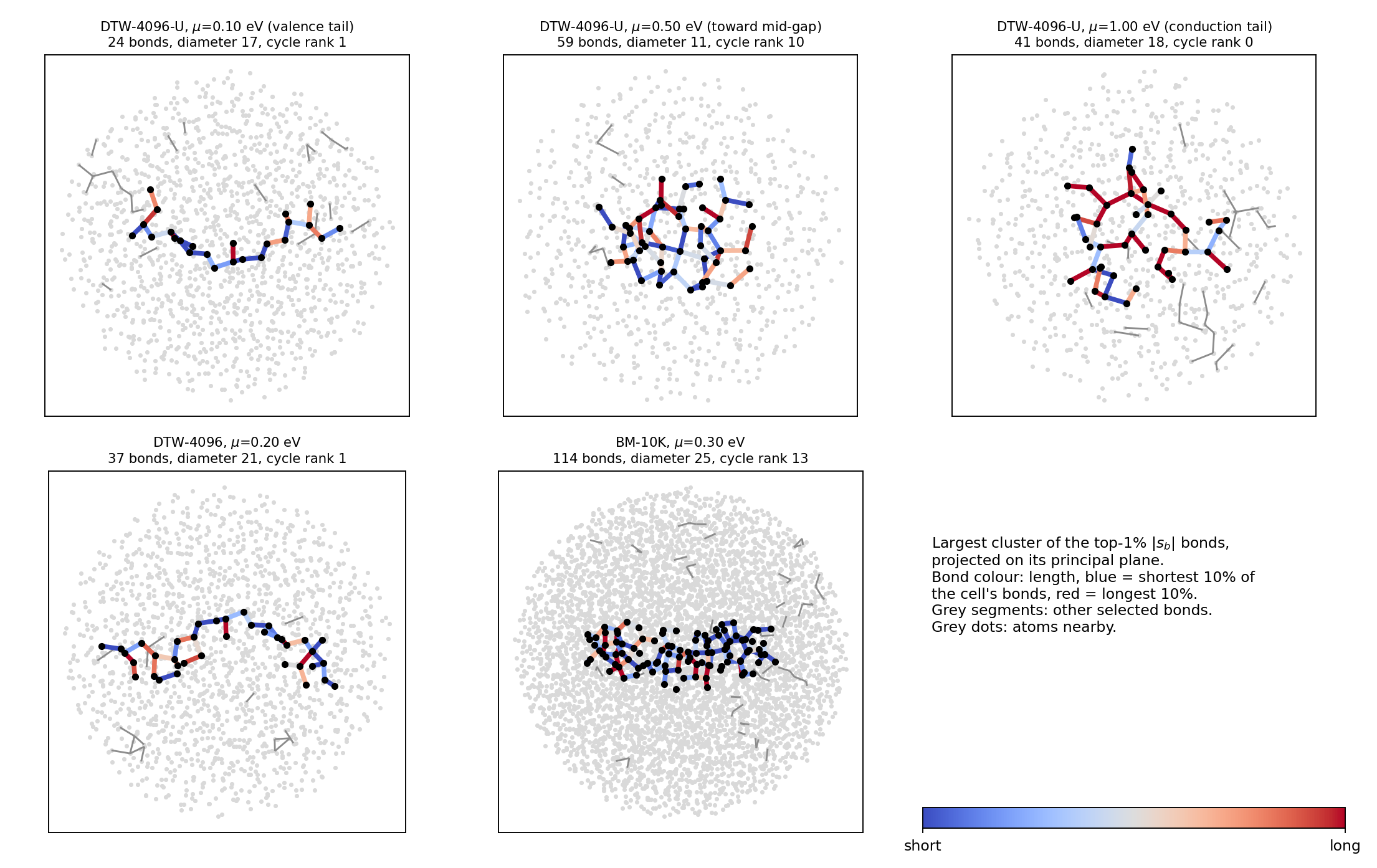}
\caption{Largest cluster of the top-1\% $|\sbond|$ bonds, projected on its principal plane. Bond colour gives length (blue: shortest 10\% of the bonds of the network; red: longest 10\%). Grey segments are other selected bonds and grey dots nearby atoms. The valence-tail cluster is a filament, the cluster toward mid-gap is branched and looped, and the conduction-tail cluster consists mostly of long bonds.}
\label{fig:gallery}
\end{figure*}

What do the clusters look like? The shape changes with $\mu$ (Fig.~\ref{fig:topology}). At the valence-tail peaks the largest cluster is a filament. Its diameter, the longest shortest path within the cluster~\cite{West2001,Diestel2025}, is 17--21 bond steps against 4--5 for the reference (diameter excess $z_{\mathrm{dia}}=10.0$--$15.9$), and its cycle rank, the number of independent cycles~\cite{West2001,Diestel2025}, is 0 or 1. This holds at 0.075--0.125~eV in DTW-4096-U, at 0.2~eV in DTW-4096, and at 0.15~eV in BM-10K. Deeper in the gap the clusters acquire loops: the cycle rank is 10 at 0.5~eV in DTW-4096-U (cycle-rank excess $z_{\mathrm{cyc}}=8.0\pm0.2$) and 13 at 0.3~eV in BM-10K ($z_{\mathrm{cyc}}=15.3\pm0.6$). Near the conduction edge ($\mu=1.0$~eV) the clusters of the two 4096-atom networks are filaments again, with no loops and $z_{\mathrm{dia}}=4.7$--$5.5$, but they are small (14 and 41 bonds). Over 30 maps the median cycle rank is 1.0 against 0.1 for the LDOS-weighted reference, and the median diameter 9.5 bond steps against 5.0. Figure~\ref{fig:gallery} shows examples.

Do short and long bonds matter? Take the shortest and longest 10\% of the first-neighbour bonds of each network. At valence-tail energies 20--40\% of the selected bonds are short, against 10\% for uniform selection and 12--27\% for the LDOS-weighted reference. At $\mu\geq1.0$~eV, 47--83\% are long, against 29--47\% for the reference. Near mid-gap there is no preference. This is the pattern of the Urbach tails, with valence-tail states on short bonds and conduction-tail states on long bonds~\cite{Dong1998BandTail,Fedders1998,Pan2008,Drabold2011Urbach}; the filaments lie in the regions of those tail states, and Sec.~IV~H ties them to individual tail eigenstates. Much of the pattern comes from the LDOS, since the reference already carries part of it. The clusters are not pure filaments of short or long bonds: the largest valence-tail clusters are 24--35\% short bonds and the conduction-side clusters 39--50\% long bonds. Matching the reference to the bond-length distribution leaves the excess unchanged ($z_{\mathrm{LDOS}}=19.8\pm0.9$ becomes $19.7\pm0.9$ in DTW-4096 at 0.2~eV, and $5.4\pm0.1$ becomes $7.4\pm0.4$ in DTW-4096-U at 1.0~eV), so bond length does not produce the clustering.

\subsection{Conduction channels and electron eigenstates}\label{subsec:channels}

\begin{figure*}[!t]
\includegraphics[width=0.8\textwidth]{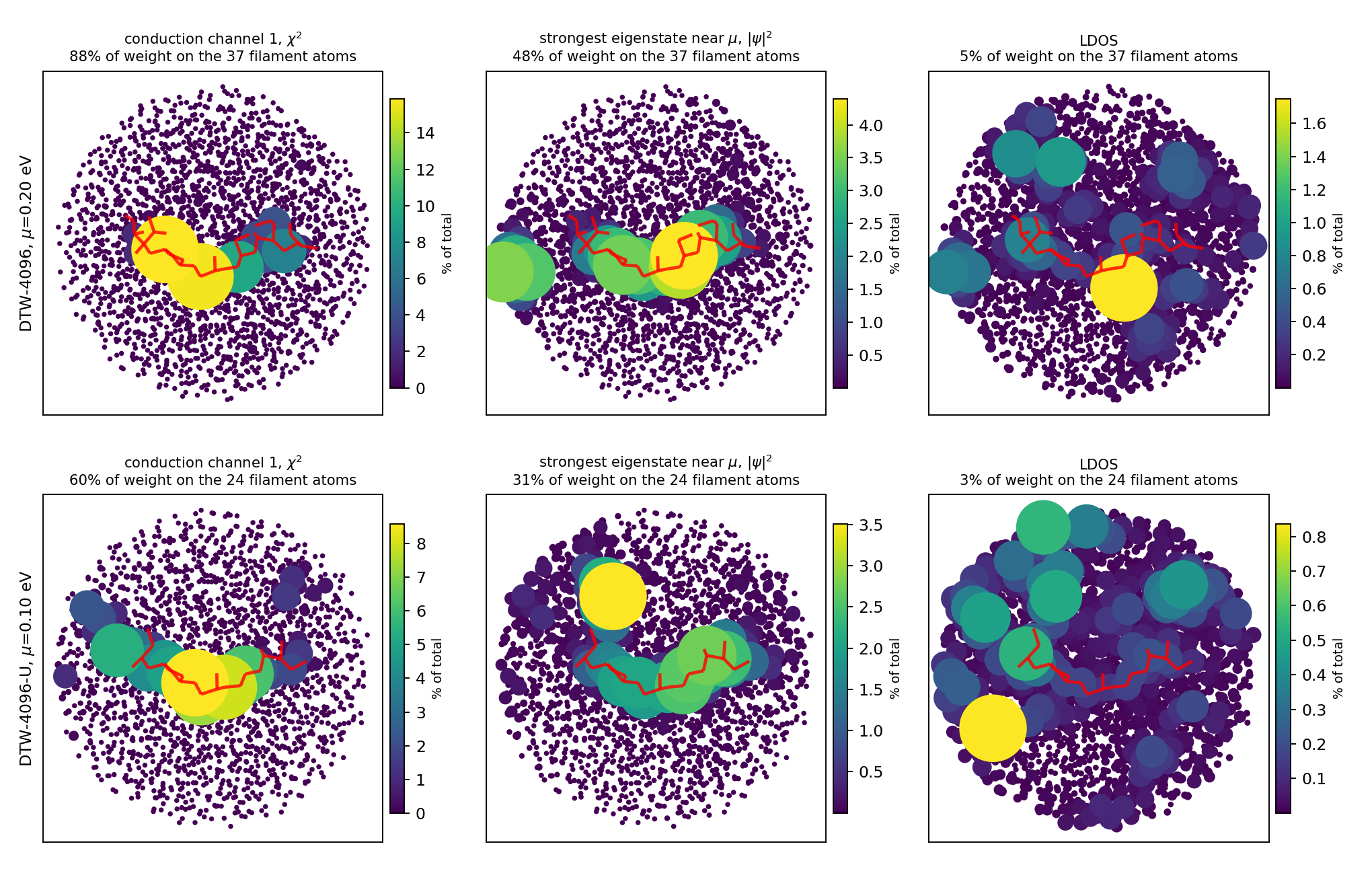}
\caption{The filament carries the leading conduction channel and a Hamiltonian eigenstate. Atoms near the largest cluster of the top-1\% $|\sbond|$ bonds (red; DTW-4096 at 0.20~eV, 37 atoms, and DTW-4096-U at 0.10~eV, 24 atoms), projected on the principal plane of the cluster; marker size and colour give each atom's share of the quantity named above the panel. Left: leading channel of the positive semidefinite $\Gamma$. Centre: the eigenstate with the largest weight on the cluster near $\mu$. Right: LDOS.}
\label{fig:states}
\end{figure*}

Which electron eigenstates make the filaments? Subedi \textit{et al.}\ diagonalized the conduction matrix $\Gamma$, whose elements $\Gamma_{xy}$ are the atom-pair contributions to $\sigma$, and found that a few eigenvectors carry the conduction path~\cite{Subedi2020SPC_2}. We repeat this in atom blocks. Appendix~\ref{app:gamma} defines $\Gamma_{xy}$. It is positive semidefinite, its elements sum to $\sigma$, and its row sums are the SPC. The conductivity is a sum of channel contributions $\Lambda_k(\mathbf 1\cdot\chi_k)^2\geq0$ (cumulative curves in Fig.~S2). Figure~\ref{fig:states} shows the result.

\begin{table*}[!t]
\caption{Filaments (largest cluster of the top-1\% $|\sbond|$ bonds) against conduction channels and Hamiltonian eigenstates. $z_A$, $z_D$: cluster excess against the LDOS-weighted and transport-window references (mean of ten sets of 400 selections; s.d.\ in Tables~S9 and~S10). $N_{50}$: channels for 50\% of $\sigma$. $w_{\mathrm{ch}}$: largest filament weight among channels 1--3. Best and nearest: excess of the Jaccard overlap $J$ over 400 LDOS-weighted random clusters, in control standard deviations, for the eigenstate with the largest filament weight and for the eigenstate nearest $\mu$.}
\label{tab:filaments}
\footnotesize
\begin{ruledtabular}
\begin{tabular}{lccccccc}
Case & Fil.\ atoms & $z_A$ & $z_D$ & $N_{50}$ & $w_{\mathrm{ch}}$ & Best $J$ exc. & Nearest $J$ exc.\\
\hline
DTW-4096, 0.10~eV & 7 & 1.3 & $-$2.1 & 17 & 0.11 & 0.6 & $-$0.6\\
DTW-4096, 0.20~eV & 37 & 19.8 & 11.6 & 6 & 0.93 & 18.5 & $-$0.5\\
DTW-4096, 1.00~eV & 15 & 6.2 & 0.5 & 26 & 0.41 & 6.5 & $-$0.5\\
DTW-4096-U, 0.10~eV & 24 & 10.2 & 7.0 & 18 & 0.60 & 9.8 & $-$0.3\\
DTW-4096-U, 0.50~eV & 50 & 9.7 & 11.2 & 2 & 0.99 & 6.8 & 0.0\\
DTW-4096-U, 1.00~eV & 42 & 5.4 & 0.2 & 18 & 0.41 & 4.5 & 1.9\\
BM-10K, 0.15~eV & 31 & 14.6 & 8.8 & 59 & 0.73 & 14.9 & $-$0.5\\
BM-10K, 0.20~eV & 38 & 16.6 & 9.6 & 27 & 0.44 & 14.7 & $-$0.4\\
BM-10K, 0.30~eV & 102 & 48.5 & 35.8 & 4 & 0.96 & 41.6 & $-$0.2\\
\end{tabular}
\end{ruledtabular}
\end{table*}

The hierarchy is gradual. In the six 4096-atom cases of Table~S10, 50\% of $\sigma$ comes from 2--26 channels out of 4096, 90\% from 11--324, and 99\% from 53--1276. The leading channel carries 8--27\%.

Do the channels coincide with the filaments? In DTW-4096 at 0.2~eV they do. The three leading channels place 88, 91, and 93\% of their weight on the 37 atoms of the filament, against 0.9\% for random atoms and 5\% for LDOS-weighted atoms, and together they carry 35\% of $\sigma$. All three are built chiefly from one Hamiltonian eigenstate, 0.017~eV below $\mu$, which puts 48\% of its weight on the filament ($z=9.7\pm0.2$ against LDOS-weighted random sets) and carries 55\% of the filament's LDOS weight. The filament is not a region of high LDOS: its LDOS weight is below that of random LDOS-weighted sets ($z=-1.8\pm0.1$). A particular state selects it. The other eight cases, including the three in the 10\,000-atom network, are in Tables~\ref{tab:filaments} and~S10; in BM-10K 50\% of $\sigma$ needs 4--59 channels ($N_{50}$ in Table~\ref{tab:filaments}).

Counting over the nine cases (Table~\ref{tab:filaments}): in the eight cases with a significant excess against the LDOS reference ($z_A>3$), one of the first three channels puts 0.41--0.99 of its weight on the filament, against 0.002--0.012 for random atoms. Six of these cases are also significant against the transport-window reference ($z_D>3$); the two that are not are the conduction-side cases at 1.0~eV.

Does one eigenstate explain the filament? We score each first-neighbour bond by $|\sum_{a\in x,b\in y}E_{ak}H_{ab}E_{bk}|$ for a state $k$, keep the same number of bonds $n_b$ as in the sensitivity selection, and compare this set with the bonds of the filament by the Jaccard overlap $J$ (Tables~\ref{tab:filaments} and~S9). The state with the largest weight on the filament is chosen for that weight, so a raw $J$ cannot be read alone. We therefore repeated the whole procedure for 400 LDOS-weighted random clusters of the same size, each against its own bonds, and report the excess over the control in units of its standard deviation. In the eight cases with a filament, the best state overlaps the filament's bonds at $J=0.15$--$0.52$, against control means of $0.03$--$0.32$; the excess is $4.5$--$41.6$ standard deviations. In DTW-4096 at 0.1~eV, where there is no filament, it is 0.6. The best state lies 17--115~meV from $\mu$. The eigenstate nearest $\mu$ gives $J\leq0.11$ with an excess of at most 1.9 (Sec.~S8).

A single state does not account for the filament. A conduction channel is an exact, non-negative part of $\sigma$, and in DTW-4096 at 0.2~eV the three leading channels and the filament coincide; we call that filament a candidate conduction pathway. We have not shown that the filaments carry most of the current: half of $\sigma$ needs 2--59 channels, and the conduction-side filaments do not survive the transport-window reference. The bonds the best state selects form a larger cluster with loops (cycle rank 4--41) than the filament (0--13). The reference also matters. Weighting random bonds by the transport window, $\rho'_i\rho'_j$ with $\rho'(x)=\sum_in_i(x)[-f'(\varepsilon_i)]$, lowers the excess in most cases and removes it at 1.0~eV in both 4096-atom networks (Table~\ref{tab:filaments}, $z_D$). The conduction-side clusters are not significant against a reference that follows the transport window.

The diagonal $\Gamma_{xx}$ is non-negative and, by the argument of Appendix~\ref{app:gamma}, gauge invariant. In six cases it correlates with KuboMap (Spearman 0.92--0.95), with the LDOS (0.92--0.97), and with $|\se|$ (0.76--0.87), and weakly with the SPC (0.39--0.49 in the four cases tested). It recovers 0.77--0.94 of the top-5\% sensitivity, less than KuboMap (0.88--0.97) in every case. Its sum is not $\sigma$ and it is not a partition (gauge test in Sec.~S8).

\subsection{Correlations between electronic and thermal maps}

We compare the electronic maps with the site-projected thermal conductivity (SPTC; $\zeta_i$) and the thermal mass-sensitivity map $\kappa_{m,i}=-2m_i\,\partial\kappa/\partial m_i$ of the companion study~\cite{Drabold2026SPTC}, which is positive when a small increase of mass lowers the thermal conductivity, using matched atoms in the same structures\footnote{The structures denoted R2--R4 here correspond to R1--R3, respectively, in the companion thermal study~\cite{Drabold2026SPTC}.}. In DTW-4096 at $\mu\leq0.3$~eV the electronic tail-weight maps are anticorrelated with $\zeta_i$ ($-0.3$ to $-0.4$) and correlated with $\kappa_{m,i}$ ($+0.2$ to $+0.35$). The ACE-relaxed DTW-512 gives the same anticorrelation with $\zeta_i$ ($-0.26$ to $-0.40$ at $\mu\leq0.55$~eV) but only a weak correlation with $\kappa_{m,i}$ (at the site level at most $+0.16$). This fits electronic tail weight sitting on soft, large-amplitude sites, which carry little heat and have positive mass sensitivity. Once each map's dependence on the vibrational amplitude is removed, little electronic--thermal correlation remains near the valence tail ($|r|\leq0.12$ in DTW-4096 at $\mu\leq0.6$~eV): both maps largely reflect local softness, and the correlations are not evidence of direct coupling between electronic and thermal transport. On the conduction side ($\mu\geq1.0$~eV) the partial correlation with $\zeta_i$ stays at $+0.35$ to $+0.47$ in DTW-4096.

In the paracrystalline models R3 and R4 (R2 and R3 of Ref.~\cite{Drabold2026SPTC}) the correlations have the opposite sign deep in the valence band ($\mu=0.05$--0.15~eV, 0.4--0.7~eV below the HOMO): $\zeta_i$ correlates at $+0.1$ to $+0.4$ and $\kappa_{m,i}$ at $-0.1$ to $-0.2$. At tail energies they have the same sign as in DTW-4096, and R2 shows no sign change. The reversal would fit electronic tail weight concentrated in crystalline grain interiors, which carry more heat per atom and have negative mass sensitivity in R3 and R4 (R2 keeps positive mass sensitivity there). The signs of the electronic on-site sensitivity and of the thermal mass sensitivity are not systematically related: regions that respond in both channels need not respond in the same direction (Sec.~S8).

\section{Scope and limitations}

The calculations use TB and DFT localized-orbital Hamiltonians at $\Gamma$, with sensitivities evaluated about fixed reference geometries. The velocity operator neglects intra-atomic dipole matrix elements~\cite{Graf1995,Pedersen2001} and adopts a displacement convention for the long-range elements of the L\"owdin-orthogonalized Hamiltonian. Suppressing elements whose separations approach half the cell length changes $\sigma$ by up to 19\%, although the maps remain highly correlated: at least 0.97 for $\se$ and KuboMap and 0.94 for SPC.

Spectral and basis choices add uncertainty. Tail conductivities vary by factors of 3--9 over $\eta=0.05$--$0.2~\mathrm{eV}$, while top-5\% map overlaps stay at 0.5--0.8. Comparisons between Hamiltonians near defect levels depend on an uncertain energy alignment. The principal DFT comparisons use a double-$\zeta$ basis; the single-$\zeta$ test described above changed $\sigma$ by 6--10\% in one cell, and basis convergence was not otherwise tested. The smallest eigenvalue of the overlap matrix is about $10^{-3}$, so L\"owdin orthogonalization may be sensitive to numerical and basis-set details.

All empirical results are for silicon, and the topology results are for three large networks. Each diagnostic answers a narrow question. Sensitivity recovery uses $|\se|$ and so does not test whether a perturbation raises or lowers $\sigma$. The gauge tests sample rotations and do not give the worst-case variation. Cluster excess depends on structure, energy, the selected bond fraction, the broadening, and temperature, and small cells have too few bonds to resolve extended connectivity. These finite models do not establish a localization transition or unique conducting pathways. The filament results rest mainly on the ACE-relaxed DTW-4096 and BM-10K and on DTW-4096-U, which is not at an ACE minimum. Several smaller cells were also used as supplied; relaxing them leaves the aggregate statistics unchanged but changes the atom-resolved maps and which runs show significant clustering (Table~S12).

The electronic--thermal comparison is atom-resolved, and the bond-level analysis is preliminary. Position sensitivities can guide infinitesimal displacements in the models studied, but designing a conductivity needs finite-perturbation tests that include structural relaxation, changes in electronic structure, and the physical realizability of the change.

\section{Conclusion and Outlook}

Mapping conductivity and finding where to change it are different problems. Partitions are not unique. SPC assigns current coupling to atoms; it is gauge invariant but signed. KuboMap assigns overlap between state pairs; it is non-negative but can depend on the eigenvector basis. For a given model and perturbation, the conductivity sensitivity is a well-defined response, and its bond decomposition contains SPC as one of three terms.

KuboMap follows the LDOS closely (median Spearman correlation 0.97). The sensitivity correlates with KuboMap at about 0.8. The local structural descriptors we tested explain at most 18\% of the variance of the local density and 9\% of the sensitivity variance, so the maps contain information that these descriptors miss. The sensitivity is the exact first-order response to on-site shifts, not to arbitrary structural changes: no tested map reliably predicts the response to removing an atom in the defective supercells, and in DTW-512 success depends on the chemical potential. Atom-level sensitivity is also structurally fragile: relaxing a cell by about 0.15~\AA{} rearranges it more than it rearranges KuboMap or the LDOS (median Spearman correlation between original and relaxed maps about 0.5 for $|\se|$; Table~S12).

How much KuboMap depends on the eigenvector basis varies with energy and structure, from negligible in some small-supercell band tails to relative differences of 0.07--0.3 elsewhere. Spectral binning restores invariance within fixed subspaces and removes the crystalline artefact, but it adds a dependence on the bin boundaries and does not make the map more stable under physical perturbations. In the thermal snapshots the mean conductivity is 2.2--2.9 times the relaxed value, with relative standard deviations of 16--24\%, which exceeds the gauge-related variations. KuboMap results should be reported with their spatial and spectral conventions and their relation to the LDOS. When the aim is the response to a specified local change, the sensitivity is the right quantity.

The most sensitive bonds cluster, and we asked two questions about the clusters. What are they? In the large networks they are tree-like filaments of 24 to 38 atoms in the valence tail, with short bonds enriched, looped networks toward mid-gap, and small filaments of long bonds near the conduction edge. The excess appears only when $\eta$ spans about five or more level spacings ($\eta/\Delta\gtrsim5$; Table~S11), so a filament is built from a group of nearby tail states rather than from one state, consistent with Sec.~IV~H. In thermal snapshots at 300 K they are largely lost ($z_{\mathrm{LDOS}}$ falls from $19.8\pm0.9$ to $4.2\pm3.3$ at 0.2~eV in DTW-4096). Are they tied to the electronic states and to transport? Partly. In the eight cases with a filament, the eigenstate with the most weight on it, a tail state within 115~meV of $\mu$, overlaps its bonds 4.5 to 41.6 control standard deviations above random clusters; the eigenstate nearest $\mu$ does not (at most 1.9). The overlap is small in absolute terms in most cases ($J=0.15$--$0.52$ against control means of $0.03$--$0.32$). One of the first three conduction channels puts 0.41--0.99 of its weight on the filament in the same cases. No single state or channel explains it: the bonds the state selects form a larger cluster with loops, and half of $\sigma$ needs 4--59 channels in the 10\,000-atom network. Six of the eight cases remain significant against a reference that follows the transport window; the two conduction-side cases at 1.0~eV do not. These results come from nine cases in three networks.

A pilot scaling of one filament's bonds (Sec.~S8) separated the filament from LDOS-matched controls but not from equally sensitive scattered bonds, and the response was nonlinear; cutting the filament lowered $\sigma$ while cutting the bonds of each control group raised it on average. A systematic test is the next step, followed by finite, physically realizable changes with structural relaxation, and their transfer across Hamiltonians and basis sets. Metals, where the states are delocalized and we expect no filaments, are the natural next test. Larger systems are needed to learn how sensitive-bond clustering scales with size. Matched electronic and thermal calculations could show whether a local modification can tune both transports.

\begin{acknowledgments}
We thank Prof.\ M. N. Kozicki for collaborations and for posing the question that initiated this work. We acknowledge support from the U.S. National Science Foundation (Grant No.~MRI-2320493) and the U.S. Office of Naval Research (Grant No.~N00014-23-1-2773). C.U. acknowledges support from the Laboratory Directed Research and Development program of Los Alamos National Laboratory under the Director's Postdoctoral Fellowship Program (Project No.~20240877PRD4). Los Alamos National Laboratory is operated by Triad National Security, LLC, for the National Nuclear Security Administration of the U.S. Department of Energy (Contract No.~89233218CNA000001). Some of the analysis scripts used in this manuscript were assisted by Claude Code (Anthropic), an AI system, in accordance with the APS policy on the use of AI tools; the authors take full responsibility for the content.
\end{acknowledgments}

\section*{Data availability}
The sparse spectral code used for the windowed calculations is the open-source \texttt{sdos} package of Ref. \cite{Drabold2026sdos}, available at \url{https://github.com/Drabold/sdos} and archived on Zenodo (concept DOI \href{https://doi.org/10.5281/zenodo.24023874}{10.5281/zenodo.24023874}). The structures, sensitivity maps, analysis scripts and their outputs supporting this study are archived on Zenodo (doi:10.5281/zenodo.24352348).

\appendix
\renewcommand{\thetable}{\thesection.\arabic{table}}
\makeatletter
\@addtoreset{table}{section}
\makeatother

\section{First variation of the Kubo--Greenwood functional}
\label{app:gradient}

We derive the matrix gradient $G=\partial\sigma/\partial H$ defined by
\begin{equation}
 \delta\sigma=\sum_{ab}G_{ab}\,\delta H_{ab}=\operatorname{Tr}(G^{\mathsf T}\delta H)
\label{eq:app_graddef}
\end{equation}
for every symmetric perturbation $\delta H$.  This derivation makes explicit the three channels of response: the direct change of the velocity operator, shifts of eigenvalues, and rotations of eigenvectors.

Write the perturbation in the instantaneous eigenbasis as
\begin{equation}
 h=E^{\mathsf T}\delta H E.
\label{eq:app_h}
\end{equation}
For a non-degenerate spectrum, first-order perturbation theory gives
\begin{equation}
 \delta\epsilon_i=h_{ii},\qquad
 \delta E_i=\sum_{k\ne i}E_k\frac{h_{ki}}{\epsilon_i-\epsilon_k}.
\label{eq:app_pt}
\end{equation}
The variation of the current matrix has a direct term and two eigenvector-rotation terms,
\begin{align}
 \delta P^\alpha_{ij}={}&\bigl[E^{\mathsf T}(\delta H\circ R^\alpha)E\bigr]_{ij}\nonumber\\
 &+\sum_{k\ne i}\frac{h_{ki}}{\epsilon_i-\epsilon_k}P^\alpha_{kj}
 +\sum_{k\ne j}P^\alpha_{ik}\frac{h_{kj}}{\epsilon_j-\epsilon_k}.
\label{eq:app_dP}
\end{align}
The kernel varies only through the eigenvalues,
\begin{equation}
 \delta W_{ij}=\partial_1w(\epsilon_i,\epsilon_j)h_{ii}
 +\partial_2w(\epsilon_i,\epsilon_j)h_{jj}.
\label{eq:app_dW}
\end{equation}
Substitution of Eqs.~\eqref{eq:app_dP} and \eqref{eq:app_dW} into Eq.~\eqref{eq:sigma}, followed by use of $w(u,v)=w(v,u)$ and $P_{ij}=-P_{ji}$, yields
\begin{equation}
 G=E\widetilde G E^{\mathsf T}+G^{(A)},
\label{eq:app_gradfinal}
\end{equation}
where the two terms are already symmetric.  The direct, or velocity-operator, contribution is
\begin{equation}
\begin{aligned}
 G^{(A)}_{ab}&=\frac{2C}{3}\sum_\alpha R^\alpha_{ab}B^\alpha_{ab},\\
 B^\alpha&=E(W\circ P^\alpha)E^{\mathsf T}.
\label{eq:app_GA}
\end{aligned}
\end{equation}
Both $R^\alpha$ and $B^\alpha$ are antisymmetric, so their Hadamard product in Eq.~\eqref{eq:app_GA} is symmetric.

The spectral contribution is most conveniently written in the eigenbasis.  Its diagonal elements are
\begin{equation}
 \widetilde G_{ii}=\frac{2C}{3}\sum_{\alpha,j}(P^\alpha_{ij})^2\,
 \partial_1w(\epsilon_i,\epsilon_j),
\label{eq:app_Gdiag}
\end{equation}
and for $i\ne k$,
\begin{equation}
\begin{aligned}
 \widetilde G_{ik}&=\frac{N_{ik}}{\epsilon_i-\epsilon_k},\\
 N_{ik}&=\frac{2C}{3}\sum_{\alpha,j}P^\alpha_{ij}P^\alpha_{kj}
 \left[W_{ij}-W_{kj}\right].
\label{eq:app_Goffdiag}
\end{aligned}
\end{equation}
Although each factor on the right side of Eq.~\eqref{eq:app_Goffdiag} changes sign under $i\leftrightarrow k$, their ratio does not: $\widetilde G_{ik}=\widetilde G_{ki}$.  Equation~\eqref{eq:app_gradfinal} is therefore the symmetric gradient appropriate for a symmetric Hamiltonian.  It is the electronic analogue of the dynamical-matrix gradient in the thermal case \cite{Drabold2026SPTC}: the first term is the spectral response and the second is the explicit response of the current operator.

\subsection{Near-degenerate and exactly degenerate levels}
\label{app:degeneracy}

Equation~\eqref{eq:app_Goffdiag} should not be evaluated by direct division when $|\epsilon_i-\epsilon_k|$ is comparable to numerical precision.  Its finite divided-difference limit is
\begin{equation}
 \widetilde G_{ik}\ \longrightarrow\ \frac{2C}{3}\sum_{\alpha,j}P^\alpha_{ij}P^\alpha_{kj}\,
 \partial_1w(\bar\epsilon,\epsilon_j),
\label{eq:app_deglimit}
\end{equation}
where $\bar\epsilon=\tfrac{1}{2}\left(\epsilon_i+\epsilon_k\right)$. In implementation we replace Eq.~\eqref{eq:app_Goffdiag} by Eq.~\eqref{eq:app_deglimit} when the level separation falls below a specified relative tolerance.  The result is insensitive to that tolerance when it is chosen well below the physical broadening $\eta$.

Within an exactly degenerate eigenspace, individual eigenvectors are not unique. The first-order eigenvalue splittings are obtained by diagonalizing the projected perturbation $P\,\delta H\,P$ within that eigenspace \cite{Kato1995}. Derivatives of smooth matrix functions can instead be formulated using the Fr\'echet derivative, independently of the arbitrary eigenvector choice \cite{Higham2008functions}. Equations~\eqref{eq:app_Gdiag} and~\eqref{eq:app_deglimit} give the diagonal and degenerate-subspace limits, in an eigenbasis of the unperturbed Hamiltonian, of this Fr\'echet derivative, which enters the gradient calculation. For the smooth, broadened conductivity considered here, both
$\sigma$ and its full gradient $G=\partial\sigma/\partial H$, expressed in the fixed orbital basis, are unchanged by rotations of eigenvectors within an exactly degenerate subspace. Quantities constructed
from individual state probabilities $n_i(x)$ need not share this invariance.

\section{Exact global sum rules}
\label{app:sumrules}

The first sum rule follows from a uniform on-site shift.  If $H\mapsto H+\delta I$, every eigenvalue shifts by $\delta$, while all energy differences and eigenvectors are unchanged.  Consequently,
\begin{equation}
 \sigma(H+\delta I,\mu,\eta,T)=\sigma(H,\mu-\delta,\eta,T).
\label{eq:app_uniformshift}
\end{equation}
Differentiating at $\delta=0$ and using $\sum_x\Pi_x=I$ [Eq. \eqref{eq:n_i}] gives
\begin{equation}
 \sum_x\se(x)=\sum_aG_{aa}=-\partial_\mu\sigma.
\label{eq:app_sumruleone}
\end{equation}
This identity is a stringent check on the diagonal of the gradient but does not by itself test the spatial arrangement of the response.

The second rule follows from a simultaneous energy scaling.  Under
\begin{equation}
 (H,\mu,\eta,k_BT)\mapsto\lambda(H,\mu,\eta,k_BT),
\label{eq:app_scaling}
\end{equation}
the eigenvectors and Fermi occupations are unchanged, $A^\alpha$ and $P^\alpha$ acquire one factor of $\lambda$, and $W$ acquires $\lambda^{-2}$: one factor from the difference quotient and one from the normalized Gaussian.  Each term $W(P^\alpha)^2$ is therefore homogeneous of degree zero, and $\sigma$ is invariant.  Euler's theorem gives
\begin{equation}
 \sum_{ab}H_{ab}G_{ab}+\mu\partial_\mu\sigma+
 \eta\partial_\eta\sigma+T\partial_T\sigma=0.
\label{eq:app_sumruletwo}
\end{equation}
Separating diagonal and off-diagonal Hamiltonian elements yields the form in Eq. \eqref{eq:sumrules}.  The temperature derivative in Eq.~\eqref{eq:app_sumruletwo} acts through the Fermi occupations at fixed Hamiltonian, chemical potential, and broadening.

\section{Local parameter derivatives and the SPC decomposition}
\label{app:localderivatives}

The perturbation $\delta H=\delta\varepsilon_x\Pi_x$ represents an atomwise on-site shift.  Inserting it into Eq.~\eqref{eq:app_graddef} gives
\begin{equation}
 \frac{\partial\sigma}{\partial\varepsilon_x}=\operatorname{Tr}(G\Pi_x)=\sum_{a\in x}G_{aa}=\se(x).
\label{eq:app_onsite}
\end{equation}
This definition applies to a common shift of all orbitals on atom $x$.  Orbital-resolved derivatives follow by replacing $\Pi_x$ with a projector onto a specified orbital channel.

For an unordered bond $x<y$, we scale both off-diagonal blocks by $\lambda_{xy}$ while holding all other blocks fixed:
\begin{equation}
 H_{xy}\mapsto\lambda_{xy}H_{xy},\qquad
 H_{yx}\mapsto\lambda_{xy}H_{yx}.
\label{eq:app_bondscale}
\end{equation}
Since $H$ and $G$ are symmetric,
\begin{equation}
 \sbond(x,y)=\frac{\partial\sigma}{\partial\ln\lambda_{xy}}
 =2\sum_{a\in x,b\in y}G_{ab}H_{ab}.
\label{eq:app_bondsens}
\end{equation}
The factor of ``2'' accounts for the two matrix blocks and avoids double counting only when the bond is treated as unordered.

To state the relation to SPC precisely, we define the unordered-bond contribution
\begin{equation}
 \sigma_{\SPC}(x,y)=\frac{C}{3}\sum_\alpha\sum_{a\in x,b\in y}
 \left(A^\alpha_{ab}B^\alpha_{ab}+A^\alpha_{ba}B^\alpha_{ba}\right).
\label{eq:app_bondspc}
\end{equation}
Because $A^\alpha$ and $B^\alpha$ are antisymmetric, each pair of terms in Eq.~\eqref{eq:app_bondspc} is equal.  Contracting $G^{(A)}$ in Eq.~\eqref{eq:app_GA} with the scaled bond gives exactly
\begin{equation}
 2\sum_{a\in x,b\in y}G^{(A)}_{ab}H_{ab}=2\sigma_{\SPC}(x,y).
\label{eq:app_explicitbond}
\end{equation}
The remaining contraction of $E\widetilde GE^{\mathsf T}$ is the spectral response.  Splitting it into the diagonal and off-diagonal components of $\widetilde G$ defines
\begin{align}
 s_b^{\rm eig}(x,y)&=2\sum_{a\in x,b\in y}
 \left[E\operatorname{diag}(\widetilde G)E^{\mathsf T}\right]_{ab}H_{ab},\\
 s_b^{\rm vec}(x,y)&=2\sum_{a\in x,b\in y}
 \left[E\bigl(\widetilde G-\operatorname{diag}(\widetilde G)\bigr)E^{\mathsf T}\right]_{ab}H_{ab}.
\label{eq:app_bondterms}
\end{align}
Equations~\eqref{eq:app_explicitbond}--\eqref{eq:app_bondterms} prove Eq.~\eqref{eq:bonddecomp}.  SPC is therefore not the complete bond response: it is the part arising from the explicit change in the current operator, while $s_b^{\rm eig}$ and $s_b^{\rm vec}$ describe changes in the spectrum and in the eigenvectors, respectively.

A step-by-step recipe for the dense calculation is given in Sec.~S7 of the Supplemental Material.

\section{Conduction matrix in atom blocks}\label{app:gamma}

Subedi \textit{et al.}\ introduced the conduction matrix on a real-space grid~\cite{Subedi2020SPC_2}. In the localized basis we define, for each direction $\alpha$,
\begin{equation}
 \begin{aligned}
 \xi^\alpha_{ij}(x)&=\sum_{a\in x}E_{ai}\,(A^\alpha E)_{aj},\\
 \Gamma_{xy}&=\frac{C}{3}\sum_\alpha\sum_{ij}W_{ij}\,\xi^\alpha_{ij}(x)\,\xi^\alpha_{ij}(y).
 \end{aligned}
\end{equation}
Five properties follow.

(i) Since $\sum_x\xi^\alpha_{ij}(x)=P^\alpha_{ij}$, we have $\sum_{xy}\Gamma_{xy}=\sigma$.

(ii) The row sums are the SPC. With $B^\alpha=E(W\circ P^\alpha)E^{\mathsf T}$, $\sum_bA^\alpha_{ab}B^\alpha_{ab}=\sum_{ij}E_{ai}W_{ij}P^\alpha_{ij}(A^\alpha E)_{aj}$, and summing over $a\in x$ gives $\sum_y\Gamma_{xy}=\sigma_{\mathrm{spc}}(x)$.

(iii) $\Gamma$ is positive semidefinite. For any vector $v$, $v^{\mathsf T}\Gamma v=(C/3)\sum_{\alpha ij}W_{ij}\bigl(\sum_xv_x\xi^\alpha_{ij}(x)\bigr)^2\geq0$, because $W_{ij}\geq0$ for the Fermi--Dirac distribution.

(iv) $\Gamma$ is gauge invariant. Let $D$ and $D'$ be exactly degenerate subspaces, so that $W_{ij}$ is constant on $D\times D'$. Write $M^{x}=E_D^{\mathsf T}\Pi_xA^\alpha E_{D'}$. Then $\sum_{i\in D,j\in D'}\xi_{ij}(x)\xi_{ij}(y)=\operatorname{Tr}(M^xM^{y\mathsf T})$, which is unchanged by $M\mapsto Q_D^{\mathsf T}MQ_{D'}$ for orthogonal $Q_D$ and $Q_{D'}$. The diagonal $\Gamma_{xx}$ is therefore non-negative and gauge invariant. KuboMap uses the products $n_i(x)n_j(x)$ and does not have this property. The argument holds for exact degeneracy only. It does not cover near-degenerate levels.

(v) The spectral form is $\Gamma=\sum_k\Lambda_k\chi_k\chi_k^{\mathsf T}$ with $\Lambda_k\geq0$, so that
\begin{equation}
 \sigma=\sum_k\Lambda_k(\mathbf 1\cdot\chi_k)^2,
 \qquad
 \sigma_{\mathrm{spc}}(x)=\sum_k\Lambda_k(\mathbf 1\cdot\chi_k)\,\chi_k(x),
\end{equation}
and every term in the first sum is non-negative.

In the calculations the sums run over the window states of the sparse calculation, with $W_{ij}$ below $10^{-5}$ of its maximum dropped. This reproduces $\sigma$ to 0.01--0.06\%, and the row sums reproduce the SPC map to better than $6\times10^{-5}$ of its maximum.

\section{Spectrally binned KuboMap}\label{app:binned}

For the real Hamiltonians considered here, an orthogonal rotation within an exactly degenerate eigenspace changes the eigenvectors but leaves the corresponding spectral projector unchanged. Consequently, the SPC expression \(E(W\circ P^\alpha)E^{\mathsf T}\) is invariant under such rotations because the spectral kernel assigns identical weights to equal-energy states. Its local values, however, need not be positive. In contrast, the KuboMap representation in Eq.~\eqref{eq:partitions_km} depends on individual state probabilities \(n_i(x)\), which can change under rotations within a degenerate eigenspace. Their sum over the subspace is invariant; the atomwise partition is not.

To remove this basis dependence, we construct a spectrally binned KuboMap by grouping states into energy bins \(B\) of width \(\delta\) and replacing individual-state quantities with subspace invariants. Specifically, we define
\begin{equation}
    n_B(x)=\sum_{i\in B}n_i(x)
    =[E_BE_B^{\mathsf T}]_{xx},
    \label{eq:binned_n_B}
\end{equation}
where \(E_B\) contains the eigenvectors assigned to bin \(B\), and
\begin{equation}
    K_{BB'}=
    \sum_{i\in B,j\in B'}\sum_\alpha(P^\alpha_{ij})^2
    =\sum_\alpha\|P^\alpha_{BB'}\|_{\mathrm F}^{2}.
    \label{eq:binned_K_BB}
\end{equation}
Here, \(P^\alpha_{BB'}\) denotes the corresponding block of the current-operator matrix. The binned KuboMap is then
\begin{equation}
 \rho^{\delta}_{\KM}(x)=\frac{C}{3}\sum_{BB'}W(\bar\epsilon_B,\bar\epsilon_{B'})K_{BB'}
 \frac{n_B(x)n_{B'}(x)}
 {\sum_y n_B(y)n_{B'}(y)},
\label{eq:binned_KM}
\end{equation}
where \(\bar\epsilon_B\) is the mean energy of bin \(B\).

The invariance of this construction follows directly from its subspace representation. Let $Q$ be an orthogonal rotation within a subspace of exactly degenerate eigenvectors. Under the operation \(E_B\mapsto E_BQ_B\), the projector \(E_BE_B^{\mathsf T}\), and hence \(n_B(x)\), remains unchanged. Similarly, the current-operator block transforms as
\begin{equation}
P^\alpha_{BB'}\mapsto Q_B^{\mathsf T}P^\alpha_{BB'}Q_{B'},
\end{equation}
which preserves its Frobenius norm and therefore \(K_{BB'}\). Thus, Eq.~\eqref{eq:binned_KM} is invariant under independent orthogonal rotations within each bin (verified numerically to \(6\times10^{-16}\)). The construction is non-negative, coincides with the original KuboMap when no two window states occupy the same bin, and approaches it as \(\delta\to0\) for a non-degenerate spectrum \cite{Ugwumadu2026KuboMap}.

The use of bin-mean energies introduces an approximation to the spectral kernel, so the binned map does not necessarily reproduce the original conductivity exactly. For a smooth kernel, this approximation improves as the bin width decreases relative to the relevant spectral variation scales. Changing bin boundaries changes the subspaces themselves and is not covered by the invariance operation. Rotations mixing unequal-energy states likewise do not represent gauge transformations between eigenbases of the same Hamiltonian; they instead probe sensitivity to the chosen spectral resolution.

\end{document}